\documentclass[10pt,journal,a4paper,final,oneside,twocolumn]{IEEEtran}
\usepackage{amsmath,amsfonts}
\usepackage{algorithmic}
\usepackage{algorithm}
\usepackage{array}
\usepackage{textcomp}
\usepackage{stfloats}
\usepackage{url}
\usepackage{verbatim}
\usepackage{graphicx}

\usepackage{multirow}

\usepackage{amssymb}
\usepackage{bbding}
\usepackage{setspace}
\usepackage{booktabs}
\usepackage{tabularx}
\usepackage{pifont}
\usepackage[numbers,compress]{natbib}
\usepackage{subfigure}

\usepackage{color}
\usepackage{svg}
\usepackage{placeins}
\usepackage{setspace}
\graphicspath{{}}

\begin{document}
\bstctlcite{IEEEexample:BSTcontrol}

\title{Low-cost Parallel Transmission for Dense Indoor Data Collection with LoRaWAN: Time Synchronization and Resource Allocation}

\author{Junxiao~Liu,
	Xinyu~Fan,
	Luping~Xiang,~\emph{Senior~Member,~IEEE},
	and~Kun~Yang,~\emph{Fellow,~IEEE}
	\thanks{The paper was partly funded by Natural Science Foundation of China (Grant No. 62132004, 62531008), Jiangsu Major Project on Fundamental Research (Grant No.: BK20243059), Gusu Innovation Project (Grant No.: ZXL2024360), High-Tech District of Suzhou City (Grant No.: RC2025001), and Quzhou Government (Grant No. 2024D007, 2023D005). \textit{(Corresponding author: Luping~Xiang.)}}
	\thanks{Junxiao~Liu is with the School of Information and Communication Engineering, University of Electronic Science and Technology of China, Chengdu 611731, China, e-mail: liujunxiao@std.uestc.edu.cn.}
	\thanks{Xinyu~Fan is with the Qianyuan Laboratory, Hangzhou, 310024, China, e-mail: xinyufan1682@gmail.com.}
	\thanks{Luping~Xiang and Kun~Yang are with the State Key Laboratory of Novel Software Technology, Nanjing University, Nanjing, 210008, China, Institute of Intelligent Networks and Communications (NINE) and School of Intelligent Software and Engineering, Nanjing University (Suzhou Campus), Suzhou, 215163, China, and School of Information and Communication Engineering, University of Electronic Science and Technology of China, Chengdu, China. (e-mail: luping.xiang@nju.edu.cn; kunyang@nju.edu.cn). }

}
\IEEEpubid{\begin{minipage}{\textwidth}\ \centering
		Copyright (c) 2026 IEEE. Personal use of this material is permitted. However, permission to use this material for any other purposes must be obtained from the IEEE by sending a request to pubs-permissions@ieee.org.
\end{minipage}}

	\maketitle

\begin{abstract}
	LoRaWAN is a compelling low-cost solution for large-scale indoor Internet of Things (IoT) data backhaul, owing to its strong penetration capability and low power consumption. However, its default pure ALOHA access mechanism leads to severe channel contention, substantial packet loss, and reduced throughput under dense, concurrent transmissions. To overcome this, we propose a lightweight out-of-band (OOB) synchronization scheme that integrates a time division multiple access (TDMA) mechanism into commercial LoRaWAN Class~A networks. Unlike approaches requiring gateway scheduling, frequent downlink signaling, or custom hardware, our method introduces a single low-cost node providing millisecond-level alignment via a dedicated OOB synchronization channel. End devices seamlessly access this channel by briefly retuning their existing LoRa transceivers. Consequently, the scheme imposes zero downlink overhead during the steady-state reporting phase, requires no hardware modifications to gateways or end devices, and remains fully backward-compatible. This design enables collision-free scheduled channel access within the configured nominal resource capacity, thereby improving throughput and reducing contention. Real-world experiments using an indoor positioning prototype demonstrate that the proposed TDMA-LoRaWAN architecture improves system throughput by over 30\% and reduces the packet loss rate from 25.8\% to 5.02\% in a 20-node indoor deployment. Furthermore, large-scale simulations corroborate these empirical findings, support the scalability analysis under larger network sizes, and indicate improved energy efficiency per successful packet in dense network settings. These combined results demonstrate the effectiveness of the proposed approach for dense indoor IoT data collection and indicate its practical potential under high uplink reporting demands.
\end{abstract}


	\section{Introduction}
	LoRaWAN, as a prominent low-power wide-area network (LPWAN) technology, has gained substantial traction in Internet of Things (IoT) data acquisition scenarios owing to its open protocol specification, long-range communication capability, low energy consumption, and support for massive device connectivity~\cite{1}. Benefiting from strong signal penetration characteristics, a single LoRaWAN gateway is capable of covering multiple floors and penetrating complex indoor structures commonly found in large industrial or commercial environments, thereby simplifying network deployment and reducing infrastructure cost~\cite{2,3}. Consequently, LoRaWAN has emerged as an attractive, low-cost data backhaul solution that, when integrated with short-range technologies such as Bluetooth or ultra-wideband (UWB), can effectively support dense indoor positioning systems and other high-density sensing applications, such as industrial vibration monitoring, real-time structural health monitoring~\cite{peng2025simac}, and emergency event tracking~\cite{4,5,zhou2024temporal}.

	In indoor localization environments, end devices often require high-frequency uplink reporting to maintain adequate positioning accuracy. The standard LoRaWAN medium access control (MAC) protocol employs a pure ALOHA random access scheme, allowing devices to transmit immediately upon data availability. Under high reporting frequencies, increased device densities substantially elevate the probability of packet collisions, resulting in severe packet loss. Furthermore, to conserve network resources, indoor positioning systems typically emphasize uplink transmissions while minimizing downlink usage. As a result, downlink acknowledgments are rarely enabled, and retransmission procedures are generally disabled. Consequently, collisions intrinsic to ALOHA severely limit network capacity, diminish the timeliness and reliability of positioning data, and ultimately constrain the scalability and practical applicability of LoRaWAN in dense indoor localization scenarios~\cite{6}.

\IEEEpubidadjcol 
	
	To address LoRaWAN’s constrained network capacity, numerous enhancement strategies have been proposed~\cite{7,luo2025algorithm}. For instance, adaptive resource allocation mechanisms have been proposed to compensate for node clock drift at the gateway, predict potential collisions, and proactively adjust resources to avoid them~\cite{8,yan2025uav}. Other works enhance capacity by enabling concurrent decoding of overlapping packets, leveraging temporal offsets~\cite{9} or exploiting hardware imperfections and channel state information~\cite{10} to separate collided frames. A fair frame-scheduling approach that optimizes spreading factor (SF) allocation to improve concurrent transmission performance was introduced in~\cite{11}. Meanwhile,~\cite{12} achieved network-wide synchronization via gateway downlink timestamps and adopted slotted ALOHA (S-ALOHA) in place of pure ALOHA.

	Despite these advances, existing approaches generally exhibit notable limitations. Several methods require extensive modifications to the physical layer, reducing compatibility with commercial LoRa devices. Others incur increased device-side energy consumption, hardware complexity, or deployment overhead. Additionally, schemes reliant on gateway-driven synchronization or scheduling increase downlink traffic and impose additional computational load on the gateway. As a result, these solutions often fail to simultaneously satisfy the conflicting LPWAN design goals of low power consumption, low complexity, low cost, and broad compatibility, thus limiting their practical applicability in commercial LoRaWAN deployments.

	To overcome these challenges, this work introduces a time division multiple access (TDMA)-based optimization scheme that operates within the standard LoRaWAN framework. The proposed approach partitions the uplink window into multiple TDMA frames, each subdivided into time slots determined by radio-frequency configuration parameters. The conventional LoRaWAN random channel selection procedure is replaced with a deterministic channel allocation strategy that assigns each device a fixed transmission channel. End devices transmit during their designated time slots and channels, thereby enabling collision-free scheduled access within the nominal resource capacity. The application server performs centralized management of network topology and handles the scheduling and assignment of time slots and channels.
	
	To support TDMA operation, we develop a lightweight out-of-band (OOB) synchronization mechanism that provides millisecond-level timing accuracy. Specifically, this mechanism operates on an OOB synchronization channel accessed by periodically retuning the standard LoRa transceiver, thereby achieving synchronization without imposing additional network load or requiring extra radio hardware. While maintaining minimal downlink consumption, deployment cost, and device power usage, the proposed solution optimizes network resource allocation, reduces packet collisions, and improves overall network capacity and transmission efficiency. Moreover, the approach is fully compatible with standard LoRaWAN Class~A devices, shows favorable scalability in the presented analysis, and can be incorporated into existing LoRaWAN systems with limited additional deployment overhead.
	
The main contributions of this work are summarized as follows:
\begin{itemize}
	\item A lightweight OOB synchronization mechanism is designed and implemented for LoRaWAN systems. Using only a single low-cost synchronization node, the method achieves millisecond-level timing precision while remaining fully decoupled from uplink communication, thereby eliminating additional LoRaWAN channel overhead and avoiding interference with ongoing transmissions.

	\item A unified application-server mechanism is developed to manage system time-slot and channel resources. Correspondingly, a TDMA-based MAC-layer protocol is introduced at the end device to replace ALOHA-based random access, enabling collision-free scheduled transmissions within the nominal resource capacity, while supporting controlled priority-based reuse under network overload.
	
	\item All end-device enhancements are implemented through modular function calls without modifying core uplink logic, ensuring transparency to standard LoRaWAN protocol stacks and full compatibility with Class~A devices.
	
	\item A LoRaWAN-based positioning badge and complete indoor positioning system are developed for empirical validation. Experimental results verify the effectiveness of the proposed scheme in the implemented indoor testbed and support its practical applicability in such settings.
	
	\item A simulation platform is constructed to evaluate the performance of the proposed scheme under large-scale network conditions beyond the implemented testbed.
\end{itemize}
	
	The remainder of this paper is organized as follows. Section~II reviews related research efforts. Section~III introduces the system architecture and details the proposed TDMA-LoRaWAN MAC-layer design. Section~IV discusses the synchronization, resource allocation, and collision-avoidance algorithms. Section~V presents the experimental setup and performance evaluation. Section~VI concludes the paper and outlines potential directions for future research.
	
	\section{Related Work}
	LoRaWAN, standardized by the LoRa Alliance and built atop the LoRa physical layer, is designed to enable kilometer-range wireless communication at extremely low energy consumption while supporting large-scale device connectivity through a star topology comprising end devices, gateways, a network server, and application servers.
	
	The LoRaWAN MAC layer adopts a pure ALOHA uncoordinated access mechanism, which inherently limits scalability in dense deployments. As the number of devices increases, the lack of access coordination significantly elevates the probability of simultaneous transmissions and, consequently, packet collisions. This problem becomes particularly acute in indoor positioning applications, where devices must frequently transmit uplink data. Although LoRaWAN enforces duty-cycle constraints to reduce channel congestion, these limitations reduce available uplink bandwidth and do not fundamentally improve overall network capacity. As a result, system timeliness and data reliability degrade significantly, limiting the suitability of conventional LoRaWAN for high-density, high-frequency sensing applications.
	
	Extensive research has been conducted to improve LoRaWAN network capacity and mitigate packet collisions, with efforts broadly focusing on interference cancellation, SF allocation, time synchronization, and carrier sensing. In~\cite{13}, the authors introduced mLoRa, which exploits LoRa’s physical-layer properties to perform successive interference cancellation (SIC) using hierarchical chirp decoding and fine-grained signal analysis. Similarly,~\cite{14,15,16,17} proposed a series of non-standard receiver designs capable of decoding multiple overlapping LoRa signals. The work in~\cite{18} further utilized the capture effect in conjunction with SIC techniques to recover weaker colliding packets. However, these solutions require modifications to the physical layer and reliance on specialized receivers, reducing compatibility with commercial LoRa devices and increasing deployment cost—contradicting LPWAN design principles.
	
	Other efforts leverage the quasi-orthogonality of SFs to enable parallel transmissions~\cite{19}. For example,~\cite{20} proposed a lightweight two-step scheduling approach that groups nodes based on similar transmission powers to minimize the capture effect, while gateway-side coarse scheduling assigns SFs to groups to improve throughput and fairness. However, SF selection affects communication range, data rate, and robustness to interference; optimizing collision avoidance may inadvertently degrade system performance.
	
	Carrier sensing–based approaches have also been explored~\cite{21,22}. While these methods help reduce concurrent collisions, channel sensing significantly increases device-side energy consumption~\cite{csmabackoff,24}, making them less suitable for LPWAN applications requiring low complexity.
	
	Time synchronization solutions have received increasing attention. In~\cite{25}, gateway downlink timestamps were employed to synchronize nodes and partition uplink traffic into time slots, enabling S-ALOHA to theoretically double throughput. Building upon this,~\cite{26} introduced a decoding method that compares symbols across sub-slots to reconstruct collided packets. The framework in~\cite{27} integrates S-ALOHA and TDMA, providing differentiated support for periodic and burst traffic.
	
	The works in~\cite{28,29} leverage Class~B beacon frames for periodic synchronization and scheduling. By managing uplink transmission using gateway control, these methods enhance fairness and help prevent collisions. However, periodic beaconing introduces synchronization overhead~\cite{30}, consumes downlink resources, increases deployment complexity, and remains incompatible with standard Class~A devices.
	
	Request-based scheduling is investigated in~\cite{31}, wherein nodes send synchronization requests and await allocation decisions. The sMAC framework~\cite{32} further applies a maximization algorithm to assign channels or time offsets to nodes, implementing TDMA to reduce collisions. Nevertheless, these approaches require considerable downlink traffic, leading to congestion and conflicting with LPWAN design constraints.
	
	A sector-based TDMA MAC protocol was introduced in~\cite{33,34}, in which nodes independently determine their transmission parameters based on geographical location. However, this model assumes fixed node positions, lacks clock synchronization, and is vulnerable to drift-induced reliability issues. Pre-configuring Global Positioning System (GPS) coordinates also increases system complexity.
	
	OOB synchronization methods have also been reported. In~\cite{35}, devices incorporate frequency modulation (FM) receivers to obtain timing signals from broadcast FM channels, whereas~\cite{36} proposes an On-Demand MAC protocol using a low-power auxiliary transceiver for synchronization and wake-up signaling. Although these solutions reduce gateway load and avoid downlink consumption, they require additional hardware, increasing cost and system complexity.
	
	Although significant progress has been made, many existing solutions still suffer from essential limitations. Several approaches depend on extensive physical-layer modifications, incompatible with commercial hardware. Others require frequent gateway downlink transmissions, consuming limited downlink resources and overloading network infrastructure. Carrier sensing–based solutions significantly increase device power consumption, whereas other MAC-layer redesigns rely on additional hardware, increasing deployment complexity and cost. These limitations hinder the ability to achieve an optimal balance among low power consumption, low complexity, low cost, and broad compatibility, core objectives of LPWAN design, and thus restrict practical applicability in LoRaWAN deployments.

	\section{System Architecture and Protocol Description}
	To address packet collisions in LoRaWAN networks operating under high-frequency reporting workloads and to improve uplink network capacity, this work introduces a TDMA-based optimization scheme that remains fully compatible with the standard LoRaWAN MAC layer. By achieving network-wide time synchronization and allocating uplink time slots and channels deterministically, the proposed design replaces random-access contention with scheduled access. This section details the architecture of the TDMA-LoRaWAN system and the implementation of the MAC-layer protocol.

	\subsection{System Architecture Overview}
	
	The proposed TDMA-LoRaWAN scheme is built atop the standard LoRaWAN MAC layer and thus preserves compatibility with Class~A devices and existing commercial LoRaWAN infrastructure. The system adopts the conventional LoRaWAN star topology~\cite{37}, comprising end devices, gateways, a network server, an application server, and lightweight synchronization nodes.
	\begin{figure}
		\centering
		\includegraphics[width=3.5in]{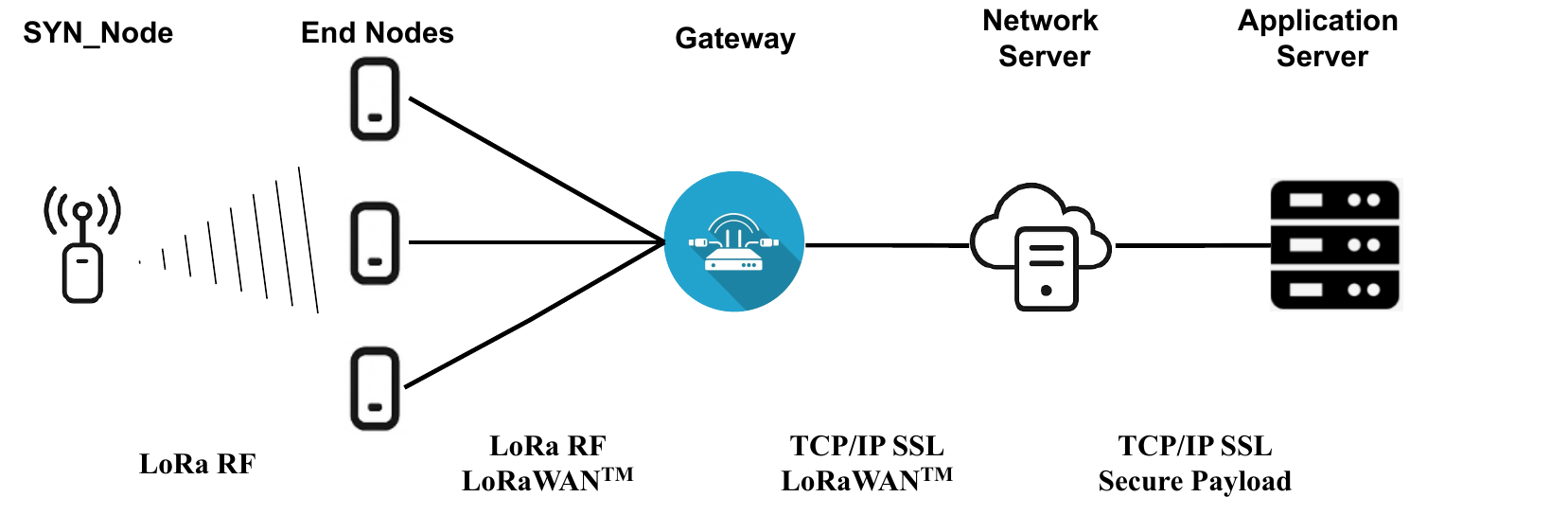}
		\caption{System network architecture.}
		\label{System_network_architecture}
	\end{figure}
	
	Gateways and network servers rely entirely on off-the-shelf LoRaWAN equipment without requiring hardware or firmware modifications. A synchronization node, implemented as a specialized LoRa end device, periodically broadcasts timing information. To adhere to low-cost and lightweight design principles, this node utilizes a high-stability temperature-compensated crystal oscillator (TCXO, $\pm 0.5$ ppm) as its time reference. This configuration supports millisecond-level synchronization precision while operating on an OOB synchronization channel to avoid interference with regular uplink transmissions.
	
	End devices execute TDMA-based uplink scheduling through enhanced MAC-layer logic that replaces the pure ALOHA mechanism. Each device adheres strictly to its allocated time slot and assigned channel, thereby avoiding random-access collisions. The application server maintains network topology information and manages the allocation of time slots and channel resources. Additional functionality can be integrated depending on application requirements. In this study, the application server also provides indoor positioning computation and visualization.

	\subsection{TDMA Implementation over LoRaWAN}
	
	The TDMA-based access protocol is implemented directly at the end-device MAC layer. Uplink resources are partitioned into time-slot and channel pairs, with each end device assigned to a unique resource block. Devices transmit exclusively within their designated block, enabling scheduled channel access without random-access contention.
	
\begin{figure}[!htbp]
		\centering
		\includegraphics[width=3in]{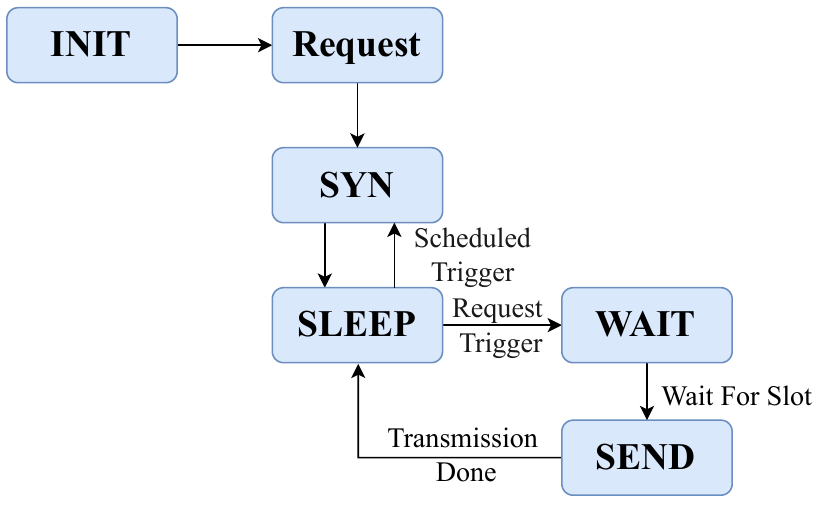}
		\caption{Finite-state machine of the TDMA-LoRaWAN protocol.}
		\label{State_Diagram}
	\end{figure}

	The MAC-layer scheduling logic is implemented as a finite-state machine, illustrated in Fig.~\ref{State_Diagram}. The state machine consists of six operational states:
	
	\begin{enumerate}
		\item INIT. The microcontroller (MCU) peripherals and LoRa radio frequency (RF) module are initialized. The device uses pre-configured access slots to issue join and access requests to the LoRaWAN network.
		\item REQ. The device transmits resource-allocation requests to the application server, seeking time-slot and channel assignments.
		\item SYNC. The device switches to the OOB synchronization channel, receives timestamp broadcasts from the synchronization node, and updates its local clock accordingly.
		\item WAIT. Prior to transmission, the device monitors the current time and waits until its assigned time slot begins.
		\item SEND. The RF module is activated and the uplink frame is transmitted.
		\item SLEEP. When no transmission tasks are pending, the device enters low-power sleep mode.
	\end{enumerate}
	
	Upon startup, the end device initializes its hardware and RF module and subsequently issues network-access requests using reserved time slots under the standard LoRaWAN procedure. After successful network association, the device submits time-slot request frames to the network server, which forwards them to the application server. The application server updates the network topology and returns downlink configuration messages that specify the allocated time slot and channel.
	
	The device then switches to the OOB synchronization channel to receive periodic timing broadcasts from the synchronization node. By calibrating its local clock, the device achieves millisecond-level timing alignment and subsequently enters a low-power mode. Periodic synchronization tasks maintain long-term timing stability without impacting LoRa uplink resources.
	
	When an uplink transmission is required, the device checks whether the current time falls within its assigned slot. If not, it remains in the WAIT state until the scheduled slot is reached. Once the uplink frame is transmitted, the device immediately returns to the SLEEP state to maintain low energy consumption.
	
	Since this study targets high-frequency uplink scenarios such as indoor positioning, unacknowledged uplink frames are employed to avoid retransmission overhead and improve system responsiveness, with the system tolerating a small degree of packet loss.
	
	\section{Proposed Optimization Algorithm}
	Building upon the system architecture described above, this section delves into the core algorithmic contributions that support scalable operation and scheduled collision avoidance within the nominal resource capacity. We elaborate on the operational logic of the proposed scheme, which comprises the OOB synchronization mechanism, the centralized resource allocation algorithm, and the hierarchical superframe scheduling method for heterogeneous traffic. These algorithms collectively support millisecond-level slot alignment and efficient spectrum utilization. For clarity and ease of reference, the key notations and symbols used throughout the protocol design and analysis are summarized in Table~\ref{tab:notations}.
	\subsection{Synchronization Scheme: Out-of-Band Time Synchronization atop LoRaWAN}
\begin{table}[!t]
	\renewcommand{\arraystretch}{1.2}
	\caption{List of Key Notations and Variables}
	\label{tab:notations}
	\centering
	\resizebox{3.4in}{!}{
		\begin{tabular}{|c|l|}
			\hline
			\textbf{Symbol} & \textbf{Description} \\
			\hline
			$T_{\text{sync}}$ & Synchronization broadcast interval \\
			\hline
			$T_{\text{to}}$ & Timeout threshold for scanning sync beacons \\
			\hline
			$T_{\text{cur}}$ & Timer tracking the active listening duration \\
			\hline
			$\Delta T$ & Calibrated local clock offset \\
			\hline
			$T_{\text{ToA}}$ & Time-on-air duration of a packet \\ 
			\hline
			$T_{\text{prop}}$ & Propagation delay \\
			\hline
			$T_{\text{irq}}$ & Interrupt handling delay on the end device \\
			\hline
			$T_{\text{encode}}, T_{\text{decode}}$ & Hardware processing delays (encoding and decoding) \\ 
			\hline
			$\Delta T_{\text{sync,max}}$ & Maximum network-wide synchronization error \\
			\hline
			$\Delta T_{\text{drift}}$ & Maximum accumulated clock drift \\
			\hline
			$\Delta T_{\text{hw,max}}$ & Maximum hardware timing jitter \\
			\hline
			$\mathbf{M}$ & Device management table maintained by the server \\
			\hline
			$\mathbf{R}$ & Time-frequency resource occupancy table \\
			\hline
			$\mathcal{C}$ & Set of candidate resource blocks for allocation \\
			\hline
			$\mathcal{A}$ & Output allocation result set \\
			\hline
			$T_{\text{last}}$ & Timestamp of the last successful activity for a device \\
			\hline
			$T_{\text{idle}}$ & Duration a device has been idle \\ 
			\hline
			$T_{\text{release}}$ & Timeout threshold for resource reclamation \\
			\hline
			$N_{\text{slots}}$ & Number of consecutive slots required by a device \\
			\hline
			$n_{\text{multi}}$ & Total number of slots occupied by multi-slot allocations \\ 
			\hline
			$P_d$ & Device priority (lower value indicates lower priority) \\
			\hline
			$N_c$ & Total number of available frequency channels \\
			\hline
			$\mathrm{Load}(C_i)$ & Normalized occupancy load of channel $C_i$ \\
			\hline
			$\rho_{\max}$ & Maximum quota ratio for multi-slot allocations \\
			\hline
			$M_{\text{super}}$ & Superframe size (total base frames, $M_{\text{super}}=2^K$) \\
			\hline
			$K$ & Maximum depth of the superframe hierarchy \\
			\hline
			$T_0$ & Duration of the baseline frame (shortest period) \\
			\hline
			$n$ & Number of time slots per baseline frame \\
			\hline
			$L_s$ & Duration of a single TDMA time slot \\
			\hline
			$T_{\text{guard}}$ & Guard time reserved within a slot \\
			\hline
			$f$ & Current frame index within a superframe \\
			\hline
			$p_i$ & Activation interval (periodicity) of device $i$ (in frames) \\
			\hline
			$g_i$ & Frame group index (offset) assigned to device $i$ \\
			\hline
			$s_i$ & Time slot index assigned to device $i$ within a frame \\
			\hline
			$k_i$ & Period exponent for device $i$ ($p_i = 2^{k_i}$) \\
			\hline
			$N_{DL}$ & Number of downlink control messages per device \\
			\hline
			$E[N_{\text{UL}}]$ & Expected number of uplink packets per session \\
			\hline
			$\eta$ & Control-to-Data Ratio (Normalized Overhead) \\
			\hline
			$T_{\text{up}}$ & Uplink reporting interval \\
			\hline
			$T_{\text{session}}$ & Average duration of a stable connection session \\
			\hline
	\end{tabular}}
\end{table}
	The adoption of pure ALOHA in LoRaWAN reflects a deliberate design trade-off between MAC-layer simplicity and system-level overhead. By accepting protocol-layer unreliability, LoRaWAN minimizes device-side complexity, reduces power consumption, and lowers deployment cost. Conventional synchronization approaches generally require continuous channel monitoring, which substantially increases energy usage. For example, Class~B devices consume 30\%--50\% more energy than Class~A devices due to periodic beacon listening~\cite{38}.

	Beyond power consumption, system resource constraints also limit the feasibility of traditional synchronization schemes. LoRaWAN’s star topology centralizes uplink processing at gateways, while downlink bandwidth remains scarce. Existing synchronization mechanisms typically require extensive downlink signaling, occupying capacity otherwise reserved for application-critical traffic. Other solutions introduce additional hardware or raise deployment costs. Given the sporadic and aperiodic nature of typical LoRaWAN traffic, the lightweight ALOHA protocol remains a pragmatic choice, effectively minimizing overhead by eliminating the need for channel sensing or scheduling~\cite{39}.
	
	While such design choices make LoRaWAN highly suitable for large-scale low-power deployments, the absence of synchronization significantly restricts its ability to support high-frequency, real-time applications. In dense indoor positioning scenarios, frequent random uplink transmissions from numerous devices generate severe collisions and drastically reduce network capacity. Achieving system-wide timing coordination is therefore essential.
	
	To this end, we propose a lightweight OOB synchronization mechanism that achieves millisecond-level precision without imposing significant additional network load, hardware complexity, or deployment cost. The scheme maintains full compatibility with the standard LoRaWAN protocol stack and requires only a single low-cost synchronization node equipped with a high-precision clock. This node periodically broadcasts compact timestamp packets (4 bytes) on an OOB synchronization channel distinct from standard LoRaWAN uplink bands, fully decoupling synchronization from data traffic. End devices retune their existing LoRa transceivers to the OOB synchronization channel to receive these beacons. To minimize channel occupancy and ensure regulatory compliance, the synchronization node transmits beacons using SF7. With a $T_{\text{ToA}}$ of 36~ms and a broadcast interval of 4~s, the resulting operational duty cycle is 0.9\%, strictly adhering to the regional industrial, scientific, and medical (ISM) band regulation limit (typically 1\%).

	\begin{figure*}[!t]
		\centering
		\includegraphics[width=6in,height=2.6in]{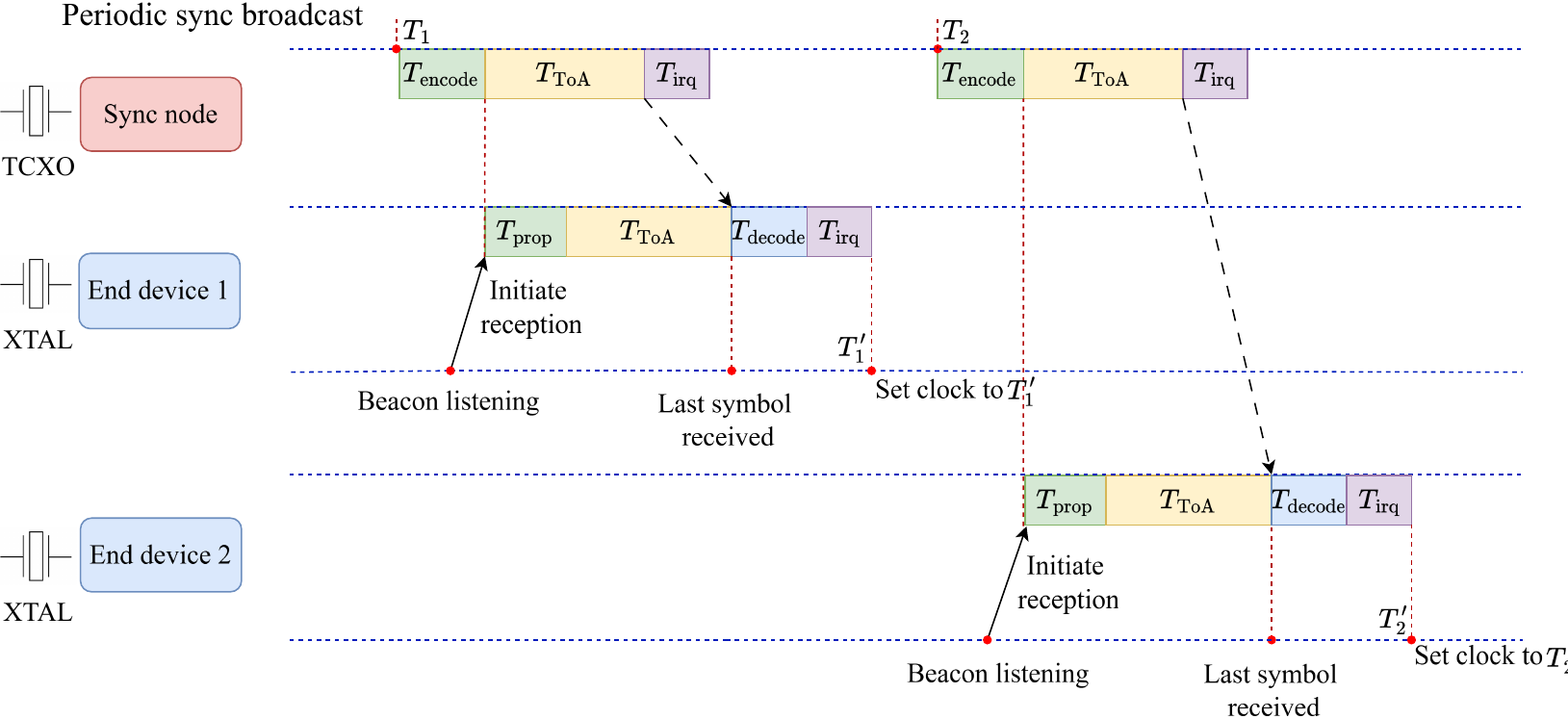}
		\caption{Synchronization process sequence diagram.}
		\label{sequence diagram}
	\end{figure*}
	
	Fig.~\ref{sequence diagram} illustrates the timing sequence of the synchronization procedure. The primary contributors to synchronization error include~\cite{40}:
	\begin{enumerate}
		\item Time-on-air ($T_{\text{ToA}}$). Transmission duration determined by RF parameters and packet size.
		\item Propagation delay ($T_{\text{prop}}$). LoRa signal propagation time, negligible in indoor applications.
		\item Interrupt handling time ($T_{\text{irq}}$). Duration required for interrupt processing on end devices.
		\item RF state transition delay. Encoding delay ($T_{\text{encode}}$) and decoding delay ($T_{\text{decode}}$), where $T_{\text{decode}}$ ranges from 176~$\mu$s to 1487~$\mu$s for SF7--SF10~\cite{40}.
	\end{enumerate}
	
	The timestamp reconstruction at the end device is therefore expressed as:
\begin{equation}
	\label{eq:1}
	\begin{aligned}
		T_1' = T_1 + T_{\mathrm{prop}} + T_{\mathrm{encode}} + T_{\mathrm{ToA}} + T_{\mathrm{decode}} + T_{\mathrm{irq}}.
	\end{aligned}
\end{equation}

	The airtime is computed using \eqref{eq:2}--\eqref{eq:5}:
\begin{equation}
	\label{eq:2}
	\begin{aligned}
		T_{\text{ToA}} &= T_{\text{preamble}} + T_{\text{payload}},
	\end{aligned}
\end{equation}
\begin{equation}
	\label{eq:3}
	\begin{aligned}
		T_{\text{payload}} &= N_{\text{payload}}\, T_{\text{sym}},
		T_{\text{sym}} &= \frac{2^{\mathrm{SF}}}{\mathrm{BW}},
	\end{aligned}
\end{equation}
\begin{equation}
	\label{eq:4}
	\begin{aligned}
		T_{\text{preamble}} &= (N_{\text{preamble}} + 4.25)\, T_{\text{sym}},
	\end{aligned}
\end{equation}
\begin{equation}
	\begin{aligned}
		N_{\text{payload}} = &\; 8 + 
		\max \Bigg(
		\left\lceil
		\frac{8\mathrm{PL} - 4\mathrm{SF} + 28 + 16\mathrm{CRC} - 20\mathrm{H}}
		{4(\mathrm{SF} - 2\mathrm{DE})}
		\right\rceil \\
		& \times (\mathrm{CR} + 4),\; 0
		\Bigg).
	\end{aligned}
	\label{eq:5}
\end{equation}

	\begin{table}[!t]
		\centering
		\caption{LoRaWAN Physical Layer Parameters for Airtime Calculation}
		\label{tab:phy_params}
		\renewcommand{\arraystretch}{1.2}
		\resizebox{\columnwidth}{!}{
		\begin{tabular}{|c|l|l|}
			\hline
			\textbf{Symbol} & \textbf{Definition} & \textbf{Configuration / Value} \\ \hline
			$\mathrm{PL}$ & Payload Length & Variable (1--255 bytes) \\ \hline
			$\mathrm{SF}$ & Spreading Factor & 7--12 \\ \hline
			$\mathrm{CRC}$ & Cyclic Redundancy Check & 1 (Enabled) \\ \hline
			$\mathrm{H}$ & Header Mode & 0 (Explicit Header) \\ \hline
			$\mathrm{DE}$ & Low Data Rate Optimization & Auto (1 if $T_{\text{sym}} > 16~\text{ms}$) \\ \hline
			$\mathrm{CR}$ & Coding Rate & 1 (Rate 4/5) \\ \hline
			$\mathrm{BW}$ & Bandwidth & 125~kHz \\ \hline
		\end{tabular}
		}
	\end{table}

	The parameters used in the equations are listed in Table~\ref{tab:phy_params}. Because LoRa airtime spans tens to hundreds of milliseconds, millisecond-level synchronization suffices for TDMA scheduling. Pursuing microsecond-level precision would introduce unnecessary overhead. After eliminating negligible terms, the timestamp estimation can be simplified to:
\begin{equation}
	\begin{aligned}
		T_1' = T_1 + T_{\text{encode}} + T_{\text{ToA}} + T_{\text{decode}}.
	\end{aligned}
	\label{eq:revised_timestamp}
\end{equation}

	In addition to synchronization error, clock drift must also be considered. Except for the TCXO-based synchronization node, end devices use low-cost crystal oscillators (20~ppm/°C), resulting in up to 60~ms drift per hour. Devices therefore periodically re-enter the synchronization state to maintain slot alignment. The synchronization algorithm is summarized in Algorithm~\ref{alg:1}.
	
\begin{algorithm}[htbp]
	\caption{Time Synchronization Algorithm for LoRaWAN End Devices}
	\label{alg:1}
	\begin{algorithmic}[1]
		
		\REQUIRE $Pkt$: broadcast packet; $T_{\text{to}}$: timeout;
		$T_{\text{sync}}$: sync period; 
		$T_{\text{cur}}$: current timer; $RF$: RF parameters
		\ENSURE $\Delta T$: calibrated timestamp
		
		\STATE Sync node broadcasts $Pkt$ periodically
		\STATE Device switches to sync channel and initializes timer
		\STATE $T_{\text{cur}} \leftarrow 0$
		
		\WHILE{$T_{\text{cur}} < T_{\text{to}}$}
		\IF{received $Pkt$}
		\STATE Extract $T_1$ from $Pkt$
		\STATE $T_{\text{ToA}} \leftarrow f(RF)$
		\STATE $\Delta T \leftarrow T_1 + T_{\text{ToA}} + T_{\text{decode}} + T_{\text{encode}}$
		\STATE Set device clock using $\Delta T$
		\STATE $T_{\text{cur}} \leftarrow 0$
		\STATE \textbf{break}
		\ELSE
		\STATE Continue listening for $Pkt$
		\ENDIF
		\ENDWHILE
		
		\IF{$T_{\text{cur}} \ge T_{\text{to}}$}
		\STATE Synchronization failed; wait for a shortened retry interval
		\STATE Re-enter synchronization mode
		\ENDIF

		\WHILE{$T_{\text{cur}} < T_{\text{sync}}$}
		\STATE Wait
		\ENDWHILE
		
		\STATE \textbf{goto} step 2
		
	\end{algorithmic}
\end{algorithm}

	Comprehensive experiments involving one synchronization node and 20 end devices show that the proposed scheme achieves an average synchronization error of 2~ms, with a maximum error of 4~ms. Although prior work reports 10~$\mu$s precision~\cite{36,40}, such solutions require substantial architectural changes and resource overhead. In contrast, our approach offers a lightweight, low-cost, and fully LoRaWAN-compliant design with practical applicability in dense indoor settings.

	To ensure resilience against temporary beacon loss or external interference, the protocol incorporates a fail-safe holdover mechanism in which the device continues operating based on its local clock for a short duration. The slot-level guard time is conservatively configured as 55~ms in the current implementation, based on the worst-case timing budget formalized in Section~IV-B. This margin protects against the aggregate timing uncertainty considered in the slot design. After a synchronization timeout, the device re-enters the synchronization state after a shortened retry interval, allowing synchronization to be restored before timing errors accumulate excessively.

\subsection{Resource Allocation}

	Each end device is assigned an uplink period corresponding to the TDMA frame length. The frame is subdivided into $n$ slots of fixed duration $L_s$. To ensure collision-free transmission despite clock drift and hardware latencies, the slot structure is defined as follows:
	\begin{equation}
		L_s = T_{\text{ToA}} + T_{\text{guard}},
	\end{equation}
	where $T_{\text{ToA}}$ denotes the time-on-air of the standard uplink packet, and $T_{\text{guard}}$ is the guard time reserved to accommodate timing uncertainties. 
	
	To guarantee collision-free transmission, the guard time $T_{\text{guard}}$ must strictly exceed the worst-case aggregate timing deviations. Considering that adjacent devices in consecutive slots may drift in opposite directions, the guard time constraint is defined as:
\begin{equation}
	\label{eq:guard_constraint}
	T_{\text{guard}} \ge 2 \cdot (\Delta T_{\text{sync,max}} + \Delta T_{\text{drift}} + \Delta T_{\text{hw,max}}),
\end{equation}
	Here, $\Delta T_{\text{sync,max}}$ denotes the baseline synchronization error, $\Delta T_{\text{drift}}$ represents the maximum accumulated clock drift, and $\Delta T_{\text{hw,max}}$ accounts for hardware jitter, while the factor of 2 addresses the bidirectional timing deviations between adjacent slots.
	
	Empirically, with a measured $\Delta T_{\text{sync,max}} \approx 4$~ms and a calculated maximum drift of $\Delta T_{\text{drift}} = 12$~ms (based on $\pm 20$~ppm accuracy over a 600-s interval), the aggregate requirement is approximately 32~ms. Consequently, our configured guard time of 55~ms provides a robust safety margin, strictly bounding the transmission within the allocated slot even under worst-case timing jitters.
	
\begin{figure}[!htbp]
		\centering

		\includegraphics[width=3.6in]{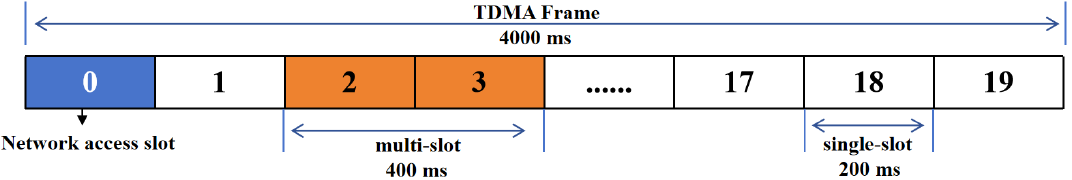}
		\caption{TDMA frame structure.}
		\label{TDMA_FRAME}

	\end{figure}

	While typical end devices share identical RF configurations, specific devices may require longer airtime. To support such heterogeneity while preserving scalability, a multi-slot mechanism is introduced, allowing devices to occupy multiple consecutive slots. As depicted in Fig.~\ref{TDMA_FRAME}, the frame structure accommodates these allocations, with Slot 0 of Channel 0 exclusively reserved for network access requests.
	
	Since LoRa signals on non-overlapping channels do not interfere, retaining random channel selection would underutilize the available channel resources. Thus, channel resources are explicitly treated as allocable entities, and each device is assigned a deterministic slot–channel pair.

	The allocation procedure is summarized in Algorithm~\ref{alg:2}. Resource allocation is centralized at the application server to guarantee deterministic operation while minimizing downlink signaling. The resource allocation algorithm relies on a device management table $\mathbf{M}$, recording device states and static priority levels $P_d$ (with lower values signifying lower criticality), alongside a real-time time-frequency occupancy map $\mathbf{R}$. Based on these inputs, a hybrid strategy is implemented, merging load-aware first-fit allocation with priority-aware admission control under network overload.

To formalize the allocation metric driving this strategy, we define the per-channel load as the normalized occupancy of channel $C_i$ within the current frame:
\begin{equation}
	\mathrm{Load}(C_i) \triangleq \frac{1}{n}\sum_{u=0}^{n-1}\mathbb{I}\!\left\{\mathbf{R}[i][u]\neq\varnothing\right\},
\end{equation}
where $n$ is the number of slots per frame and $\mathbb{I}\{\cdot\}$ is the indicator function. Under this definition, minimizing $\mathrm{Load}(C_i)$ is equivalent to selecting the least-occupied channel, thereby spreading allocations uniformly across channels and reducing localized congestion. Notably, $\mathrm{Load}(\cdot)$ is independent of payload length and only reflects spectrum occupation, ensuring that the allocation policy remains lightweight and stable. The scheduling objective is to achieve collision-free allocation within the nominal resource capacity, while preserving high-priority access and limiting excessive multi-slot occupation when the network becomes overloaded.

When feasible idle resources exist (i.e., the candidate set $\mathcal{C}\neq\varnothing$ in Algorithm~\ref{alg:2}), the scheduler chooses the candidate resource block on the least-loaded channel and allocates it exclusively to the new device. However, when no feasible idle resource exists ($\mathcal{C}=\varnothing$), the system reaches its nominal resource capacity limit. In this saturated state, the scheduler switches to a priority-based resource reuse rule. Specifically, the server allows a high-priority request to reuse a block currently occupied by a lower-priority or long-inactive device (selected based on the ranking tuple $(\mathrm{Priority},-T_{\text{idle}})$). This operation results in resource overbooking, where the specific time-frequency block is effectively shared by two devices. Consequently, while strict collision-free operation is guaranteed within the nominal resource capacity, operation beyond this limit degrades gracefully through controlled resource reuse, resulting in localized contention on shared resources. This ensures that high-priority end devices can still access the network during congestion, albeit with a risk of collision on the shared resource.

To prevent multi-slot users from monopolizing the frame under heterogeneous airtime demands, multi-slot admission is bounded by a configurable cap. Let $n_{\text{multi}}$ denote the total number of slots currently occupied by multi-slot allocations. A new multi-slot request is admitted only if
\begin{equation}
	\frac{n_{\text{multi}}}{N_c \cdot n} \le \rho_{\max},
\end{equation}
where $N_c$ is the number of channels, $n$ is the number of slots per frame, and $\rho_{\max}\in(0,1)$ is a system parameter (e.g., $\rho_{\max}=0.3$). This constraint limits excessive frame occupation by multi-slot allocations and helps balance heterogeneous airtime support against overall fairness and effective capacity.

\begin{algorithm}[!htbp]
	\caption{LoRaWAN Resource Allocation Algorithm}
	\label{alg:2}
	\begin{algorithmic}[1]
		\REQUIRE $Pkt$: uplink packet; $T_{\text{release}}$: timeout threshold; $L_s$: slot duration;
		\REQUIRE $\mathbf{M}$: device management table (DevID, $C_i$, $\{S_j\}$, $T_{\text{last}}$, $N_{\text{slots}}$, $P_d$, is\_reuse, is\_multi);
		\REQUIRE $\mathbf{R}$: resource table $\mathbf{R}[i][j] \in \{\text{DevID}\} \cup \{\varnothing\}$
		\ENSURE $\mathcal{A} = (C_i, \{S_j, \dots, S_{j+N_{\text{slots}}-1}\})$: Allocation result
		\STATE Get current time $T_{\text{now}}$
		\FOR{each device $d \in \mathbf{M}$} 
		\IF{$T_{\text{now}} - d.T_{\text{last}} > T_{\text{release}}$}
		\FOR{each $(C_i, S_j) \in d.\{S_j\}$}
		\STATE $\mathbf{R}[i][j] \leftarrow \varnothing$
		\ENDFOR
		\STATE Remove $d$ from $\mathbf{M}$
		\ENDIF
		\ENDFOR
		\IF{received $Pkt$}
		\STATE Extract (DevID, Type, SF, $L_{\text{payload}}$, is\_multi, $P_d$) from $Pkt$
		\IF{Type == report}
		\IF{DevID $\in \mathbf{M}$}
		\STATE Update $T_{\text{last}} \leftarrow T_{\text{now}}$ in $\mathbf{M}$
		\ENDIF
		\ELSIF{Type == request}
		\IF{is\_multi == false}
		\STATE $N_{\text{slots}} \leftarrow 1$
		\ELSE
		\STATE $T_{\text{ToA}} \leftarrow f(\text{SF}, L_{\text{payload}})$
		\STATE $N_{\text{slots}} \leftarrow \left\lceil \dfrac{T_{\text{ToA}}}{L_s} \right\rceil$
		\ENDIF
		\STATE $\mathcal{C} \leftarrow \varnothing$
		\FOR{each available pair $(C_i, S_j)$ in $\mathbf{R}$}
		\IF{$\mathbf{R}[i][j], \ldots, \mathbf{R}[i][j+N_{\text{slots}}-1] = \varnothing$}
		\STATE $\mathcal{C} \leftarrow \mathcal{C} \cup \{(C_i, \{S_j, \ldots, S_{j + N_{\text{slots}} - 1}\})\}$
		\ENDIF
		\ENDFOR
		\IF{$\mathcal{C} \neq \varnothing$}
		\STATE $(C_i, \{S_j\}) \leftarrow \arg\min\limits_{(C,\{S_u\}) \in \mathcal{C}} (\mathrm{Load}(C), u)$
		\STATE $\mathbf{R}[i][j \ldots j + N_{\text{slots}} - 1] \leftarrow \text{DevID}$
		\STATE is\_reuse $\leftarrow$ \textbf{false}
		\ELSE
		\STATE $(C_i, \{S_j\}) \leftarrow \arg\min\limits_{(C, S) \in \mathbf{R}} (\text{$P_d$}, -T_{\text{idle}})$
		\STATE $\mathbf{R}[i][j \ldots j + N_{\text{slots}} - 1] \leftarrow \text{DevID}$
		\STATE is\_reuse $\leftarrow$ \textbf{true}
		\ENDIF
		\STATE $T_{\text{last}} \leftarrow T_{\text{now}}$
		\STATE Update $\mathbf{M}$ with (\text{DevID}, $C_i$, $\{S_j\}$, $T_{\text{last}}$, $N_{\text{slots}}$, \text{$P_d$}, \text{is\_reuse}, \text{is\_multi})
		\STATE $\mathcal{A} \leftarrow (C_i, \{S_j, \ldots, S_{j+N_{\text{slots}}-1}\})$
		\STATE Send Downlink Allocation $\mathcal{A}$
		\ENDIF
		\ELSE
		\STATE \textbf{Wait for next packet}
		\ENDIF
	\end{algorithmic}
\end{algorithm}

In this allocation scheme, the control overhead per device is characterized by a fixed value of $N_{DL} = 2$, representing the initial Join-Accept and a one-time resource configuration message. Once the slot-channel assignment is established, the end device operates autonomously without requiring subsequent downlink updates or periodic signaling. Under dynamic network conditions, additional downlink traffic is only incurred during the discrete re-association process. Furthermore, the inactivity-based resource reclamation described in Algorithm~\ref{alg:2} is managed as a server-side background process and does not consume downlink resources.

To quantify the overhead of the allocation strategy, we define the control-to-data ratio (CDR), denoted as $\eta$, which represents the ratio of downlink control messages to successful uplink data packets. Let $T_{\text{up}}$ be the reporting interval and $T_{\text{session}}$ be the average duration of a stable connection. The expected number of uplink packets per session is
\begin{equation}
	E[N_{\text{UL}}] = \frac{T_{\text{session}}}{T_{\text{up}}}.
\end{equation}

Since each session requires a fixed $N_{\text{DL}} = 2$, the normalized control overhead $\eta$ is defined as the ratio of downlink control signaling to successful uplink data packets:
\begin{equation}
	\eta = \frac{N_{\text{DL}}}{E[N_{\text{UL}}]} = \frac{2 \cdot T_{\text{up}}}{T_{\text{session}}}.
\end{equation}

For a typical scenario with $T_{\text{up}} = 4$~s and a conservative $T_{\text{session}} = 24$~h, the resulting overhead $\eta$ is approximately $9.3 \times 10^{-5}$. This implies that $\eta \ll 1$, demonstrating that the control plane traffic remains orders of magnitude lower than the data plane throughput, thereby preserving the gateway's limited downlink duty cycle for massive data collection.

Based on the proposed resource allocation algorithm, the total network capacity is defined by the available time-frequency resource blocks. In our current configuration with 8 channels and 20 time slots per frame, the system supports up to 159 devices with deterministic, collision-free scheduled access (one slot reserved for network access). For scenarios where the number of sensors exceeds this limit, additional devices are managed through a controlled priority-based reuse mechanism, preserving prioritized access for mission-critical data even under saturated network conditions.
	
	\subsection{Superframe Structure}
	
While a uniform uplink period efficiently serves specialized applications like indoor positioning, it restricts the network's versatility and scalability in heterogeneous IoT scenarios. To address this limitation and accommodate devices with diverse reporting requirements, we propose a hybrid access mechanism based on hierarchical slot mapping. By building upon the implemented TDMA-LoRaWAN protocol, this mechanism integrates devices with various periods into a unified baseline frame structure through a systematic mapping strategy, which is detailed as follows:

\begin{algorithm}[!t]
	\caption{Superframe-Based TDMA Uplink Scheduling}
	\label{alg:3}
	\begin{algorithmic}[1]
		
		\REQUIRE 
		$M_{\text{super}}$ : base frames per superframe; 
		$n$: slots per base frame; $\mathcal{D}$: Device table $(\text{DevID}, k_i, g_i, s_i)$
		\ENSURE 
		Uplink transmission schedule
		
		\FOR{each device $d_i \in \mathcal{D}$}
		\STATE Compute activation interval: $p_i = 2^{k_i}$ frames
		
		\FOR{frame index $f = 0$ to $M_{\text{super}}-1$}
		\IF{$f \bmod p_i = g_i$}
		\STATE Schedule uplink packet in slot $s_i$ of frame $f$
		\ELSE
		\STATE Skip transmission
		\ENDIF
		\ENDFOR
		\ENDFOR
		
	\end{algorithmic}
\end{algorithm}
	
	Let $T_0$ denote the duration of the shortest device period (baseline frame). To support hierarchical scheduling, the superframe is defined to comprise $M_{\text{super}}$ baseline frames, where $M_{\text{super}} = 2^K$ for a system-defined maximum depth $K$. Consequently, the allowable device reporting interval $T$ must satisfy the dyadic constraint:
\begin{equation}
	T = T_0 \times 2^{k}, \quad k \in \{0, 1, \dots, K\}. 
	\label{eq:period_constraint}
\end{equation}
	
	Devices sharing the same period exponent $k_i$ form a group and are scheduled on the same slot index every $p_i = 2^{k_i}$ frames. Each device $d_i$ is assigned a frame offset $g_i \in [0, p_i-1]$ and a slot index $s_i$. The scheduling logic is formally described in Algorithm~\ref{alg:3}.

	By configuring the superframe depth $K$ and employing this hierarchical mapping, the TDMA-LoRaWAN scheme effectively accommodates heterogeneous periodicities while preserving collision-free slot allocation within the nominal resource capacity and maintaining scalability.
	
	\section{Evaluation and Analysis of Results}

	\subsection{Experimental Platform Setup}
	
\begin{figure*}
		\centering
		\includegraphics[width=5.9in]{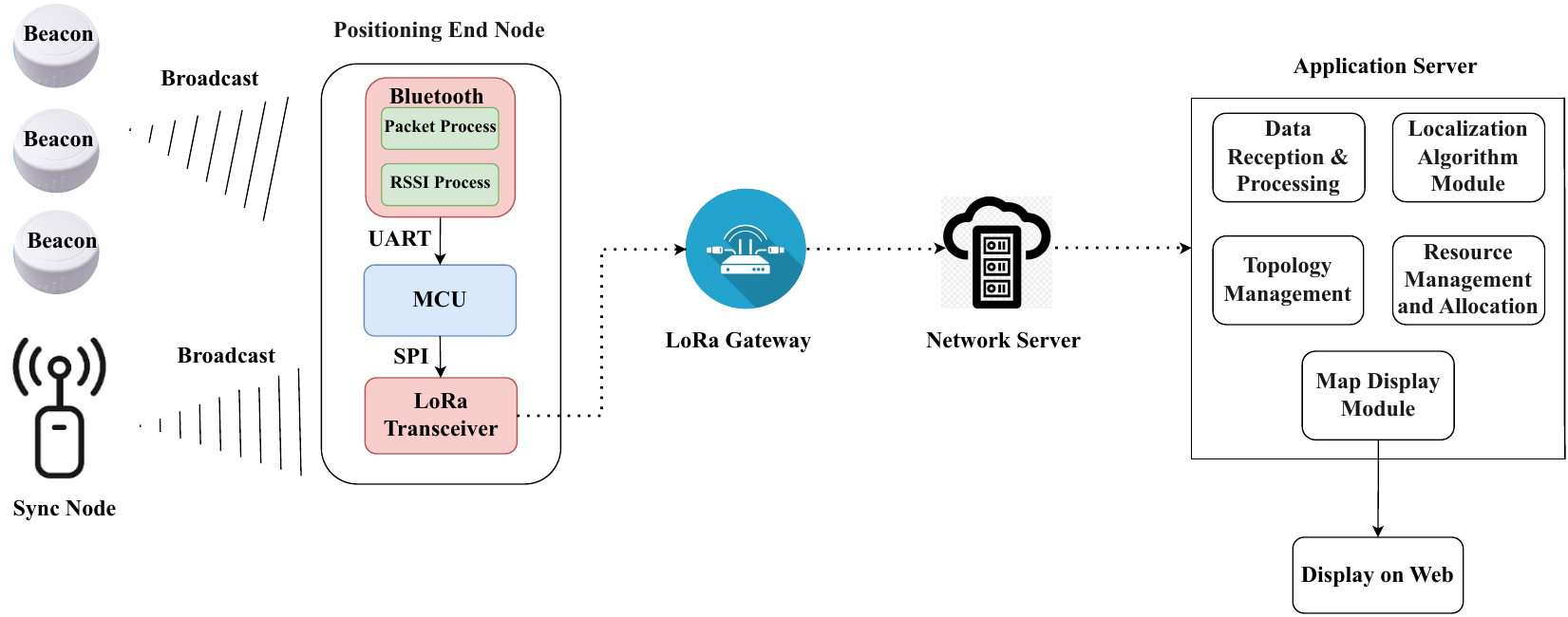}
		\caption{System architecture for the indoor positioning prototype.}
		\label{indoor_positioning_prototype_System_architecture}
	\end{figure*}

	A complete indoor positioning system employing the proposed TDMA-LoRaWAN backhaul was deployed, and system performance was evaluated under real operating conditions. As illustrated in Fig.~\ref{indoor_positioning_prototype_System_architecture}, the testbed consists of Bluetooth beacons, a synchronization node, wearable positioning tags, LoRaWAN gateways, a network server, and an application server.

	\subsubsection{End Devices}
\begin{figure}[!htbp]
		\centering
		\subfigure[]{
			\includegraphics[width=2.5in]{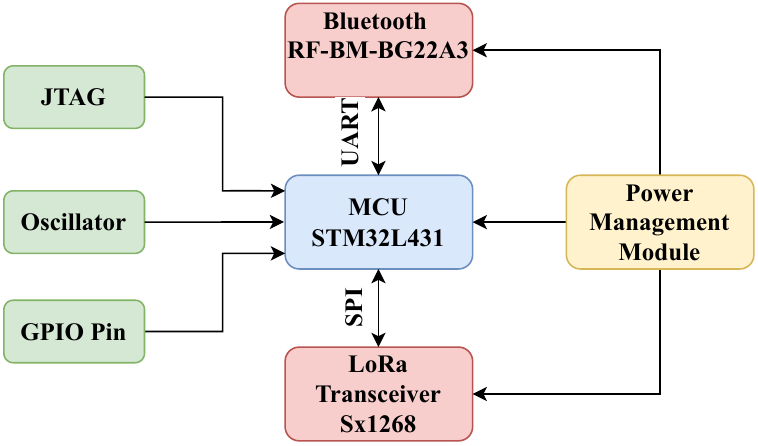}
			\label{Hardware_Diagram}
		}
		\subfigure[]{
			\includegraphics[width=2in]{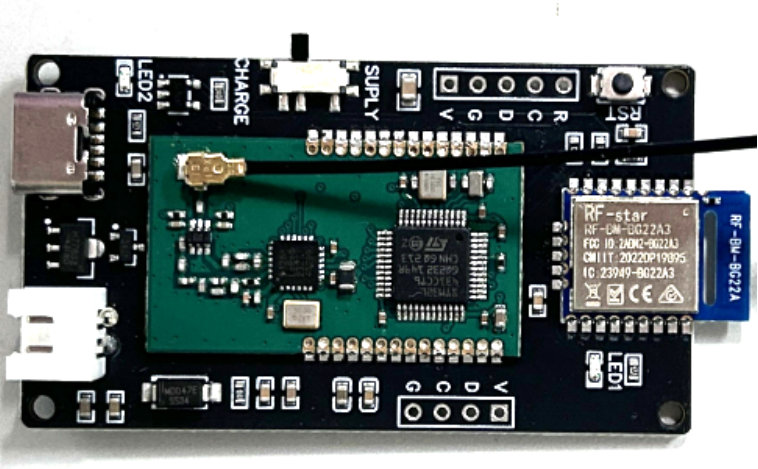}
			\label{Physical photograph}
		}
		\caption{Positioning badge hardware: (a) system block diagram and (b) prototype device.}
		
		\label{end_devices_hardware}
	\end{figure}

	Custom-designed wearable positioning badges serve as LoRaWAN end devices. As shown in Fig.~\ref{end_devices_hardware}, each badge integrates an STM32L431 MCU, an RF-BM-BG22A3 Bluetooth transceiver, an SX1268 LoRa radio, and a power management module. The device receives beacon advertisements and forwards the processed information to the server through the LoRaWAN backhaul.
	\subsubsection{Synchronization Node}
	To prioritize low cost and system simplicity, the synchronization node is implemented as a minimal hardware unit, comprising solely an STM32L431 MCU and an SX1268 LoRa transceiver. It relies on a high-stability TCXO (0.5 ppm) as its primary timing reference. In this specific experimental setup, a GPS module was temporarily integrated to serve as the ground-truth reference, enabling precise quantification of synchronization error between the sync node and end devices. Notably, this GPS module is employed strictly for performance benchmarking and is not required for the operational deployment of the proposed TDMA-LoRaWAN system.

	\subsubsection{Gateway}
	A commercial RAK7249 LoRaWAN gateway is adopted. The gateway supports Ethernet, Wi-Fi, and Long-Term Evolution (LTE) backhaul and incorporates dual SX1301 concentrators capable of configuring up to 16 uplink channels, with a maximum transmit power of 27~dBm and a receiver sensitivity of $-139$~dBm. For this evaluation, the gateway is connected to a local server via Ethernet and configured to utilize 8 uplink channels to avoid adjacent-channel interference. Combined with the 20 slots per TDMA frame, the system provides 159 resource blocks for collision-free scheduled access (with one slot reserved exclusively for network access).

	The gateway’s web-based management console offers traffic and throughput monitoring, enabling direct inspection of uplink success counts for evaluating channel contention.

	\subsubsection{Network Server and Application Server}
	A locally deployed The Things Network (TTN) instance serves as the LoRaWAN network server. TTN complies with the LoRaWAN standard and provides essential functions such as device session management, MAC-layer retransmission control, and uplink forwarding. Uplink data are delivered to the application server through the message queuing telemetry transport (MQTT) interface.
	
	The application server performs device topology management, time-slot and channel allocation, indoor positioning calculations, and map visualization.
	
	\subsubsection{Indoor Positioning Test Environment}
	Low-power Bluetooth beacons based on the EFR32BG22 platform are deployed throughout the test environment using the iBeacon protocol, with a transmit interval of 100~ms and a power level of 0~dBm. Beacon metadata are collected and stored in a cloud database.
	
\begin{figure}[!htbp]
		\centering
		\includegraphics[width=3in,height=2in]{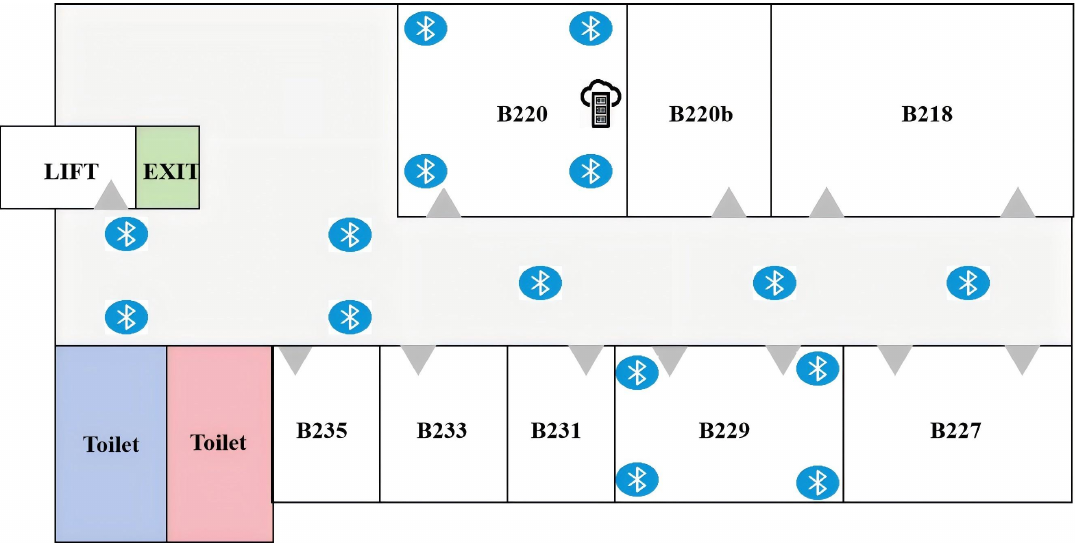}
		\caption{Deployment of Bluetooth beacons in the indoor test area.}
		\label{Deployment_of_beacons}
	\end{figure}

	Fifteen beacons are arranged in a grid pattern across the indoor space. LoRa gateways and synchronization nodes are centrally deployed, and servers operate locally. Test personnel equipped with positioning badges engage in unrestricted movement to emulate real-world usage. Deployment locations are shown in Fig.~\ref{Deployment_of_beacons}.

	The application server continuously processes uplink packets, performs multi-dimensional data fusion, and updates device positions on the indoor map in real time. The position update frequency provides an indirect measure of uplink reliability, while gateway traffic logs serve as a quantitative indicator of network capacity.

	\subsection{Test Results and Analysis}
	
	To evaluate the performance of the communication system, we conducted a comparative analysis of channel access performance between the standard LoRaWAN and the proposed TDMA-LoRaWAN protocols. Traffic monitoring data were collected from both the gateway and the network server. The gateway logs the total number of raw packets received, while the network server parses the payloads to identify unique device reports.

\begin{table}[!t]
	\renewcommand{\arraystretch}{1.15} 
	\caption{End Device Parameter Configuration}
	\label{tab2}
	\centering
	\footnotesize 
	\begin{tabular}{|l|c|}
		\hline
		\textbf{Parameters} & \textbf{Value} \\
		\hline
		Number of End Nodes & 20 \\
		\hline
		TX Periods [s] & 4 \\
		\hline
		Payload Length [Bytes] & 10 \\
		\hline
		Preamble Length [Symbols] & 8 \\
		\hline
		Time-on-Air [ms] & 145 \\
		\hline
		Spreading Factor & 9 \\
		\hline
		Bandwidth [kHz] & 125 \\
		\hline
		Coding Rate & 4/5 \\
		\hline
		Slot Length [ms] & 200 \\
		\hline
		Slots per Period & 20 \\
		\hline
	\end{tabular}
\end{table}
	The RF parameters for the end devices are configured as listed in Table~\ref{tab2}. It is important to note that the selected reporting interval of 4~s (corresponding to a duty cycle of approximately 3.6\% per device for SF9) represents a high-traffic stress test configuration. This setting was specifically chosen to emulate the stringent requirements of real-time indoor positioning, where high-frequency updates are essential for trajectory tracking. Furthermore, this saturated traffic load allows us to evaluate the limits of the protocol's collision avoidance performance under worst-case contention scenarios.
	
\begin{figure}[!htbp]
		\centering
		\includegraphics[width=3.5in]{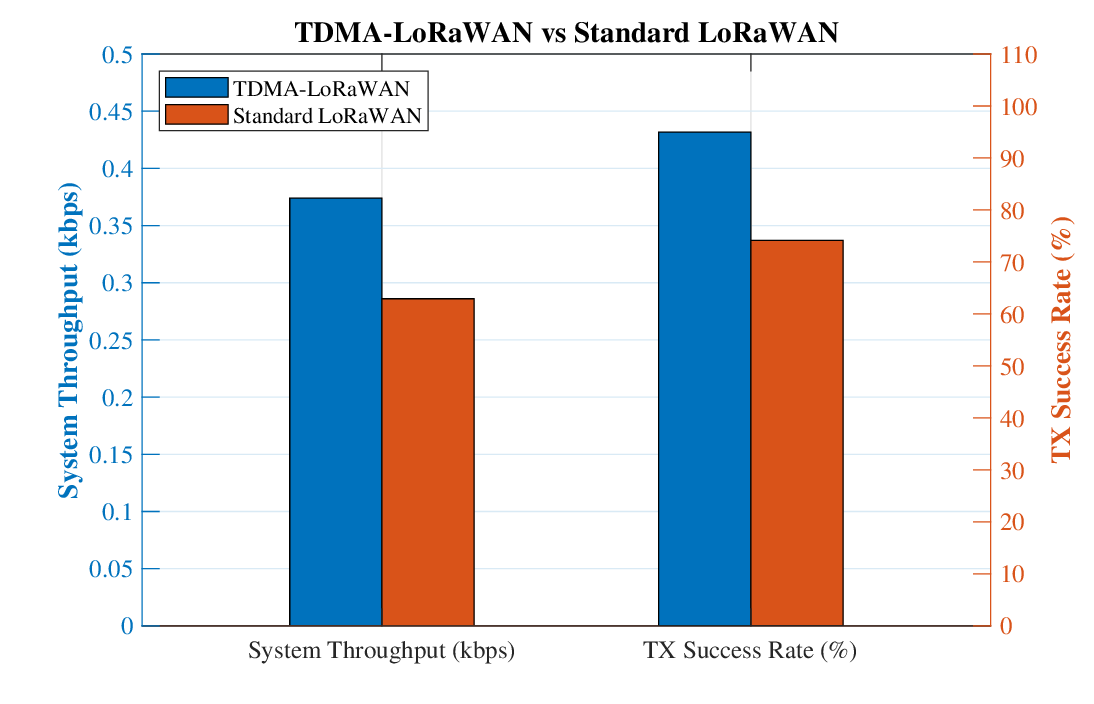}
		\caption{Experimental comparison of standard LoRaWAN and TDMA-LoRaWAN.}
		\label{Experimental_Results}
	\end{figure}

	The comparative experimental results are summarized in Table~\ref{tab3} and Fig.~\ref{Experimental_Results}. To ensure a fair comparison despite differing test durations (3100~s vs. 4050~s), we normalized the traffic metrics. As shown in Table~\ref{tab3}, the calculated packets per device per unit time (normalized load) were 0.241 and 0.246 for the Standard and TDMA scenarios, respectively. Correspondingly, the offered traffic loads were 0.386~kbps and 0.393~kbps. The negligible difference ($<$2\%) confirms that both protocols were evaluated under comparable traffic intensity.

\begin{table}[!t]
	\renewcommand{\arraystretch}{1.15} 
	\caption{Comparison Between Standard LoRaWAN and TDMA-LoRaWAN}
	\label{tab3}
	\centering
	\footnotesize 
	\begin{tabularx}{\columnwidth}{|X|c|c|}
		\hline
		\textbf{Parameter} & \textbf{TDMA} & \textbf{Standard} \\
		\hline
		Test Duration [s] & 4050 & 3100 \\
		\hline
		Normalized Load [pkt/s/node] & 0.246 & 0.241 \\
		\hline
		Offered Traffic Load [kbps] & 0.393 & 0.386 \\
		\hline
		Channel Utilization & 8.46\% & 6.49\% \\
		\hline
		Packets Sent & 19,916 & 14,968 \\
		\hline
		Packets Received & 18,917 & 11,102 \\
		\hline
		Packet Delivery Ratio (95\% CI) & 94.98\% $\pm$ 2.31\% & 74.17\% $\pm$ 3.87\% \\
		\hline
		System Throughput (kbps) & 0.374 & 0.286 \\
		\hline
	\end{tabularx}
\end{table}

	To provide a comprehensive evaluation, the statistical analysis was conducted over the entire experimental duration. Specifically, the full-duration dataset was partitioned into temporal segments to compute 95\% confidence intervals (CI), thereby explicitly incorporating overheads associated with initial network access, time synchronization, and resource allocation. To ensure that the performance differences between the protocols are attributed solely to the scheduling mechanism, downlink acknowledgments (ACKs) and retransmissions were disabled, and all physical layer parameters were maintained identical across both experimental groups.
	
	Based on this comprehensive analysis, the proposed TDMA-LoRaWAN achieved an average packet delivery ratio (PDR) of 94.98\% ($\pm 2.31\%$ CI). In comparison, standard LoRaWAN achieved 74.17\% ($\pm 3.87\%$ CI). These results confirm that deterministic scheduling effectively mitigates channel contention not only in the steady state but also during dynamic transition phases.

	In terms of spectral efficiency, TDMA-LoRaWAN achieved an effective channel utilization of 8.46\%, significantly outperforming the 6.49\% utilization of the standard protocol. Collectively, these results demonstrate that under the current test environment, the proposed scheme increases system throughput by approximately 30\% and reduces the packet loss rate by over 20 percentage points. This confirms that deterministic scheduling effectively converts offered load into successful throughput with minimal resource waste, mitigating the inherent contention limitations of LoRaWAN networks.
		
	Finally, it is crucial to contextualize these results within the system's capacity limits. With 20 devices distributed across 8 channels, the experimental system operated at approximately 12.5\% of its slot capacity. The results reveal a critical insight that, even at this moderate load, standard LoRaWAN suffers significant degradation (25.8\% loss), whereas TDMA maintains high reliability. To further evaluate system performance under full capacity and large-scale deployment scenarios, where physical testing is infeasible, the comprehensive simulations presented in the following section are employed.
		 
	 \subsection{Simulation Platform Testing}
	 Given the practical challenges of deploying hundreds of physical devices, we developed a discrete-event simulator using MATLAB to evaluate the scalability of the proposed scheme. This section outlines the simulation setup and parameter calibration, validates the model against experimental benchmarks, and analyzes key performance metrics, including data transmission success rate, system throughput, and average energy consumption.

	\subsubsection{Simulation Environment and Parameters}
	The simulation models a dense IoT deployment within a complex multi-room institutional environment (e.g., a university building) spanning an area of $100~\text{m} \times 100~\text{m}$. Despite the compact physical footprint, such environments are characterized by high-density obstacles that significantly attenuate radio signals. To accurately capture this behavior, we adopted a log-distance path loss model with a path loss exponent of $\gamma = 4.0$ and a reference loss $PL_0 = 40$~dB at 1~m~\cite{csmaparam,loggamma}:
	\begin{equation}
		PL(d) = PL_0 + 10 \cdot \gamma \cdot \log_{10}(d) + X_\sigma,
	\end{equation}
	where $X_\sigma \sim \mathcal{N}(0, \sigma^2)$ represents the shadowing component. This parameter choice ($\gamma=4.0$) characterizes the severe non-line-of-sight (NLOS) propagation environment found in buildings with hard partitions. To model the signal fluctuations induced by multipath fading and moving personnel, we incorporated a shadowing standard deviation of $\sigma = 6$~dB, alongside dynamic additive white Gaussian noise (AWGN) to simulate fluctuating thermal and background interference. Accordingly, the modeled noise floor was set to -117~dBm, derived from the theoretical thermal noise of a 125~kHz channel (-123~dBm) combined with a typical receiver noise figure of 6~dB. Based on this realistic propagation model, the physical and MAC layer parameters were rigorously calibrated to reflect industrial hardware constraints, as detailed in Table~\ref{sim_params}. Regarding the physical layer configuration, the base station is modeled as a standard 8-channel LoRaWAN gateway capable of simultaneously demodulating orthogonal signals. We differentiated capture thresholds by SF, using 6~dB for SF7 and a more conservative 8~dB for SF9. This margin accounts for implementation losses and the degradation of orthogonality in multipath-rich indoor channels.
		
	\begin{table}[!t]
		\centering
		\caption{Simulation Parameters and System Configuration}
		\label{sim_params}
		\renewcommand{\arraystretch}{1.15} 
		\footnotesize
		\begin{tabularx}{\columnwidth}{|X|c|}
			\hline
			\textbf{Parameter} & \textbf{Value}\\
			\hline
			Simulation Area & $100\,\text{m} \times 100\,\text{m}$ (Complex Indoor)\\
			\hline
			Path Loss Model & Log-distance ($\gamma=4$, $PL_0=40$dB)\\
			\hline
			Shadowing Std. Dev. ($\sigma$) & $6\,\text{dB}$\\
			\hline
			Noise Floor & $-117\,\text{dBm}$\\
			\hline
			Spreading Factor & $\text{SF7, SF9}$\\
			\hline
			Bandwidth & $125\,\text{kHz}$\\
			\hline
			Payload / Preamble & $10\,\text{Bytes}$ / $8\,\text{Symbols}$\\
			\hline
			Time-on-Air ($T_{\text{ToA}}$) & $41\,\text{ms (SF7)} / 145\,\text{ms (SF9)}$\\
			\hline
			Receiver Sensitivity & $-139\,\text{dBm}$\\
			\hline
			Capture Threshold & $6\,\text{dB (SF7)} / 8\,\text{dB (SF9)}$\\
			\hline
			Tx Power & $50\,\text{mW}$ ($17\,\text{dBm}$)\\
			\hline
			Rx / Listen Power & $10\,\text{mW}$\\
			\hline
			Reporting Interval & $4\,\text{s}$\\
			\hline
			TDMA Slots per Frame & $60\,\text{(SF7)} / 20\,\text{(SF9)}$\\
			\hline
			Synchronization Interval & $600\,\text{s}$\\
			\hline
			Synchronization Error ($\sigma_{\text{sync}}$) & $2\,\text{ms}$ (Std. Dev.)\\
			\hline
			Hardware Delay ($\sigma_{\text{hw}}$) & $3\,\text{ms}$ (Std. Dev.)\\
			\hline
			CSMA CCA Threshold & $-110\,\text{dBm}$ \\
			\hline
			CSMA Max Backoff Stages & $8$\\
			\hline
		\end{tabularx}
	\end{table}
For the MAC layer calibration, the carrier sense multiple access (CSMA) parameters were selected to balance collision avoidance against physical environmental constraints. The clear channel assessment (CCA) threshold was configured at -110~dBm, maintaining a 7~dB margin above the previously defined noise floor to ensure reliable signal detection~\cite{csmaparam}. The channel activity detection (CAD) duration was set to 2~ms for SF7 and 8~ms for SF9 to reflect the physical preamble locking time. While the theoretical sensing radius covers the deployment area, the stochastic shadowing ($\sigma=6$~dB) introduces signal variability, resulting in potential ``hidden node'' scenarios where the received signal power momentarily falls below the sensitivity threshold.

To complement this, we implemented a random backoff strategy with a contention window of $W=8$ slots~\cite{csmabackoff}. The duration of a single backoff slot was set to 30~ms to accommodate the maximum CAD duration and the radio state switching overheads. This configuration balances collision mitigation with the low-latency requirements of indoor positioning without introducing excessive protocol overhead.
	
			\begin{figure}[!htbp]
		\centering
		\subfigure[Impact of $\sigma_{\text{sync}}$ on PDR with fixed guard time.]{
			\includegraphics[width=0.8\linewidth]{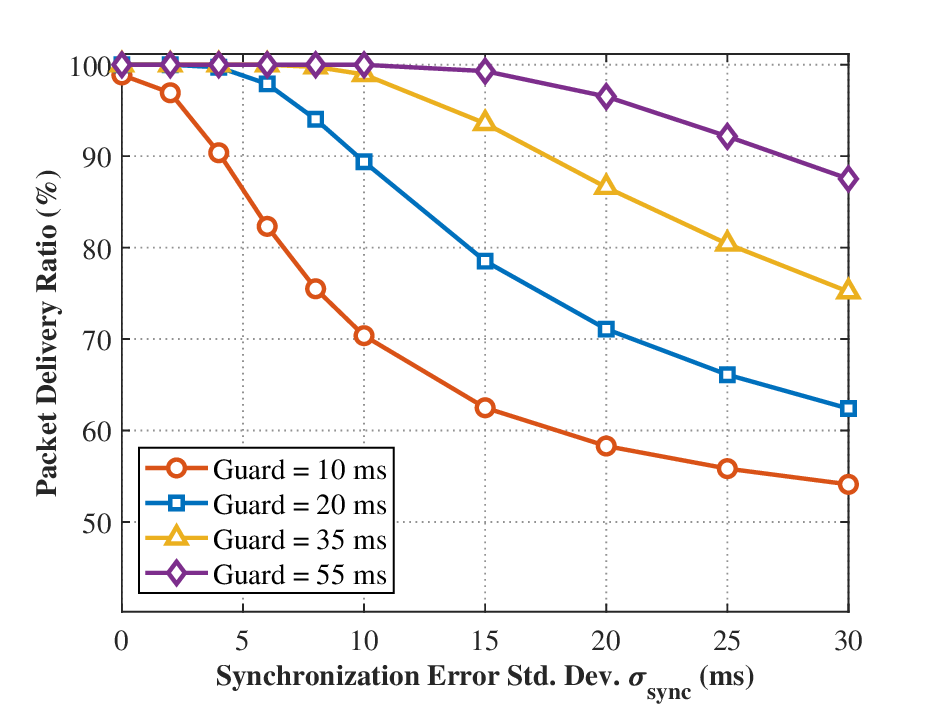}
			
		}
		\hfil
		\subfigure[Impact of $T_{\text{guard}}$ on PDR under typical and high-drift error regimes.]{
			\includegraphics[width=0.8\linewidth]{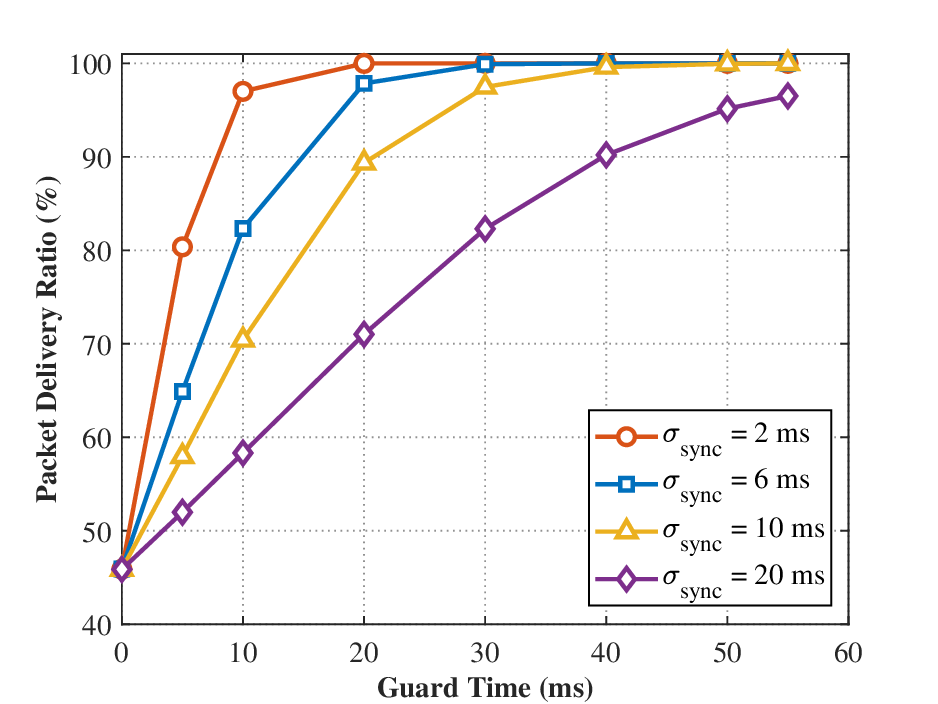}
			
		}
		\caption{Parameter sensitivity analysis regarding synchronization error and guard time.}
		\label{fig:sensitivity}
	\end{figure}			
		
	\begin{figure*}[!p]
		\centering
		\subfigure[Packet Delivery Ratio (SF7).]{
			\includegraphics[width=0.45\linewidth]{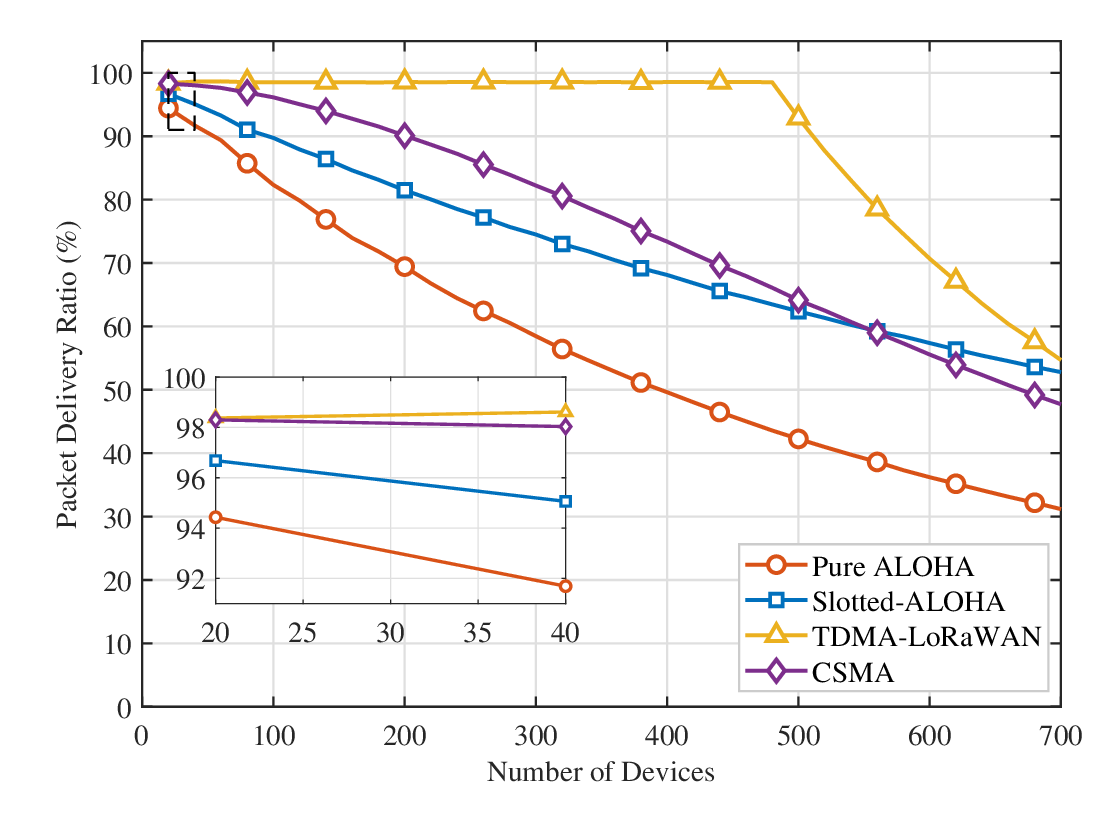} 
			\label{fig:sf7_success}
		}
		\hfil
		\subfigure[Packet Delivery Ratio (SF9).]{
			\includegraphics[width=0.45\linewidth]{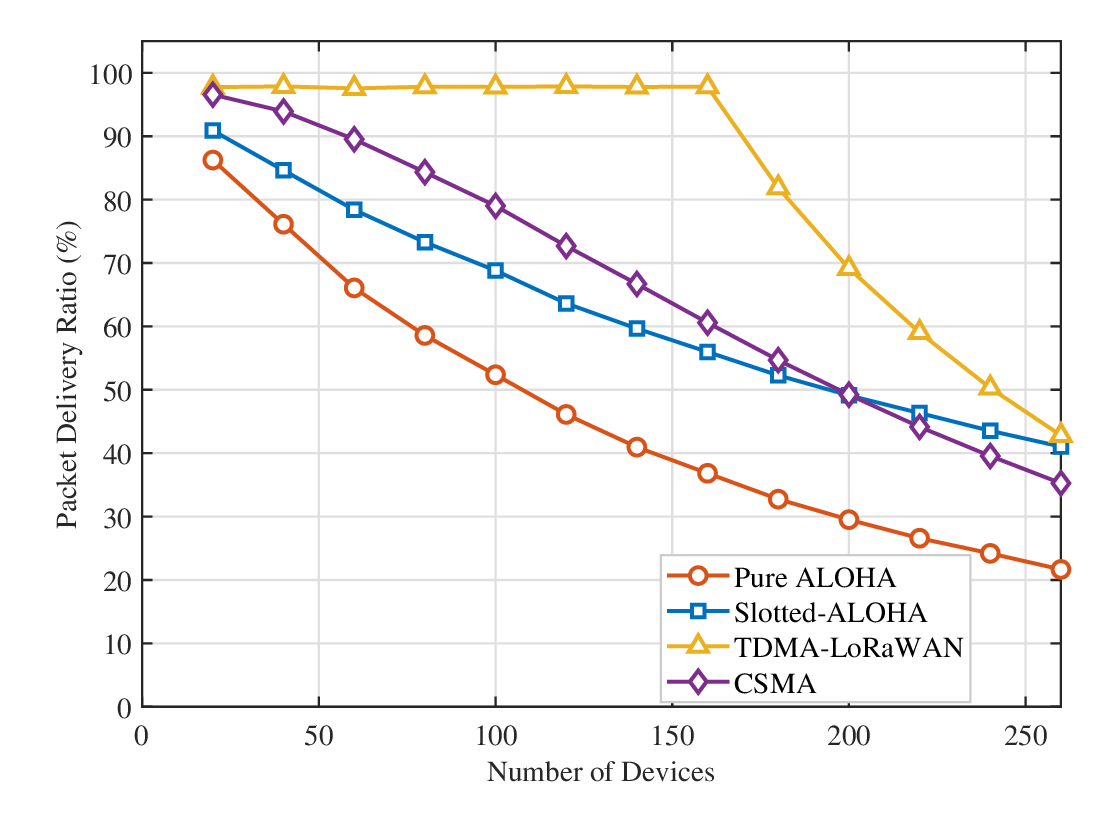} 
			\label{fig:sf9_success}
		}
		\caption{Comparison of Packet Delivery Ratio among four protocols under different SFs.}
		\label{success_comparison}
	\end{figure*}

	\subsubsection{Simulator Validation}
	To validate the simulator's fidelity, the simulation results were compared against the small-scale experimental benchmark ($N=20$, Interval = 4~s, SF9). As expected in an idealized simulation environment, the simulated performance for both protocols is consistently higher due to the absence of external interference and hardware imperfections.
	
	Specifically, the simulated TDMA-LoRaWAN PDR (97.71\%) agrees well with the experimental data (94.98\%), supporting the accuracy of the propagation modeling. For standard LoRaWAN, the simulation outperforms the testbed (86.73\% vs. 74.17\%), highlighting the impact of real-world factors such as external interference and imperfect capture effects that are simplified in the simulation model.
	
	In summary, despite the absolute performance offset resulting from environmental idealization, the simulator faithfully captures the actual PDR trends, confirming its validity for extrapolating the collision behaviors to large-scale networks.

	\subsubsection{Robustness against Timing Uncertainties}
	To evaluate the system's resilience against timing uncertainties and delineate the operational boundaries of the proposed MAC protocol, a parameter sensitivity analysis is conducted to quantify the impact of synchronization errors and guard time configurations on communication reliability. The resulting performance trends are characterized in Fig.~\ref{fig:sensitivity}.
	
	Fig.~\ref{fig:sensitivity}(a) plots the PDR against increasing synchronization errors across different guard time configurations. The results demonstrate a clear trade-off: while larger guard intervals (e.g., $T_{\text{guard}} \geq 35$~ms) sustain high reliability even under severe clock jitter, tighter guard intervals (e.g., 10~ms) exhibit a rapid performance decay as $\sigma_{\text{sync}}$ increases. This confirms that the system's tolerance to synchronization drift is strictly bounded by the allocated guard duration.
	
	Fig.~\ref{fig:sensitivity}(b) further investigates the critical boundaries by varying $T_{\text{guard}}$ under distinct error regimes. A sharp performance cliff is observed where inter-slot interference manifests. Notably, the critical threshold required to maintain collision-free scheduled transmission shifts to the right as $\sigma_{\text{sync}}$ increases (from $\sim$5~ms at $\sigma_{\text{sync}}=2$~ms to $\sim$40~ms at $\sigma_{\text{sync}}=20$~ms). This behavior empirically validates the constraint derived in \eqref{eq:guard_constraint}, illustrating that maintaining high PDR under high-drift conditions necessitates a proportional expansion of the guard interval.

\begin{figure*}[!p]
	\centering
	\subfigure[System Throughput (SF7).]{
		\includegraphics[width=0.45\linewidth]{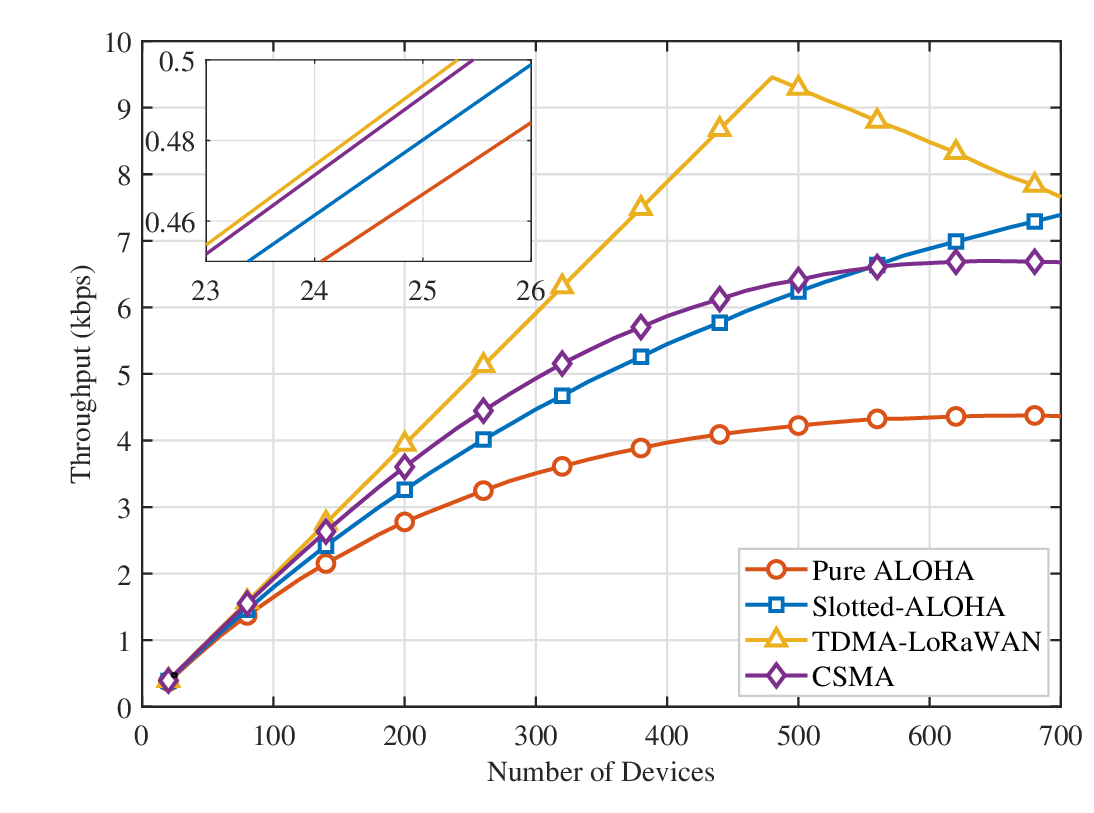} 
		\label{fig:sf7_throughput}
	}
	\hfil
	\subfigure[System Throughput (SF9).]{
		\includegraphics[width=0.45\linewidth]{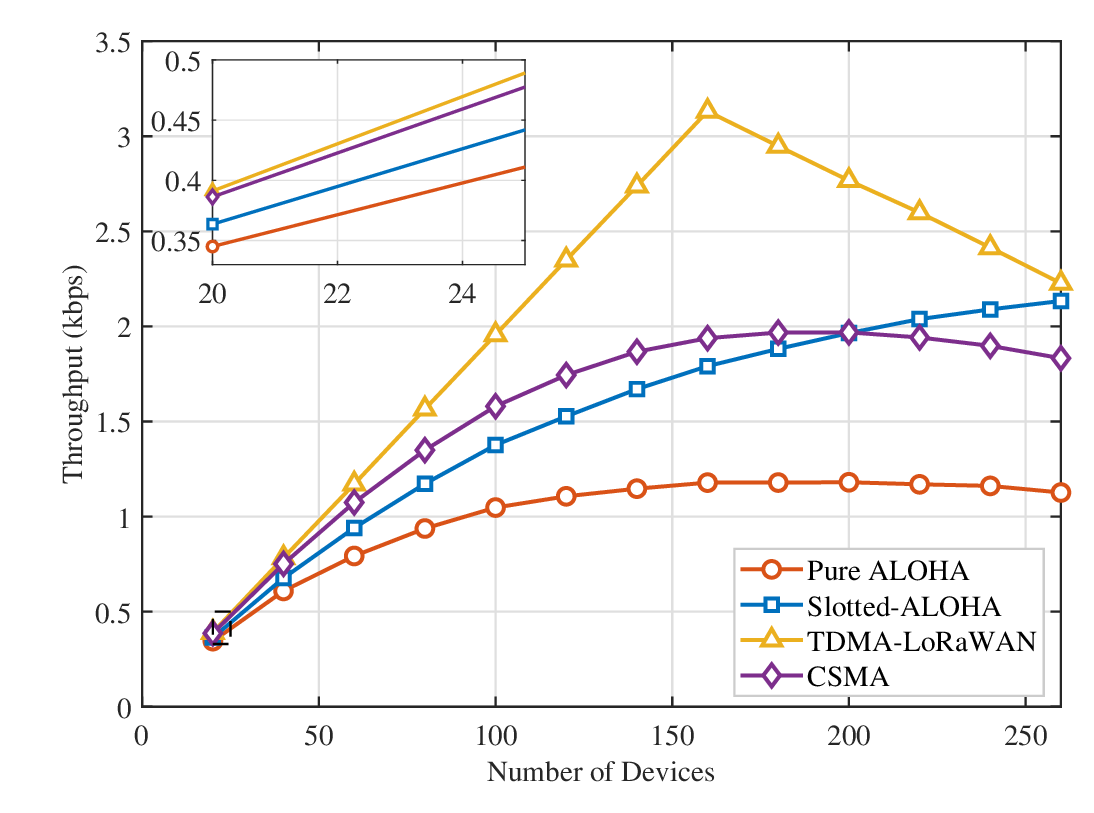} 
		\label{fig:sf9_throughput}
	}
	\caption{Throughput comparison demonstrating the scalability of the proposed TDMA scheme.}
	\label{throughput_comparison}
\end{figure*}

\begin{figure*}[!p]
	\centering
	\subfigure[Average Energy Consumption per Successful Packet (SF7).]{
		\includegraphics[width=0.45\linewidth]{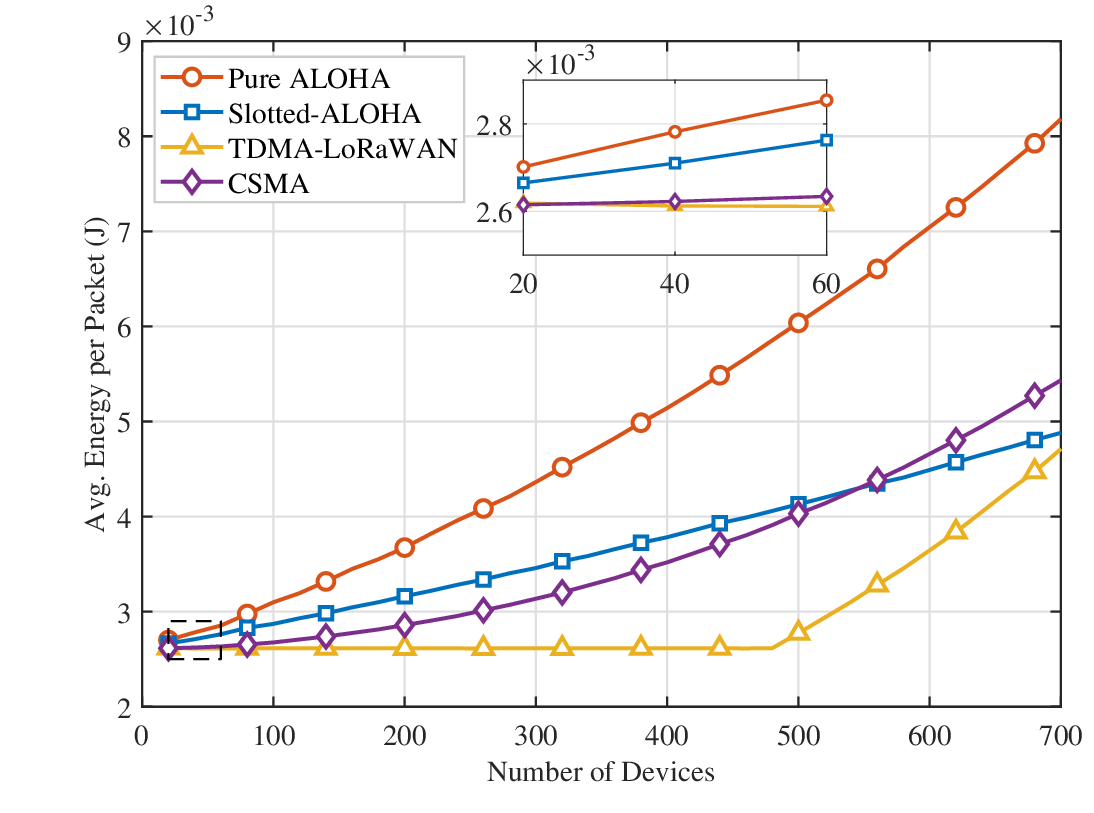}
		\label{fig:sf7_consumption}
	}
	\hfil
	\subfigure[Average Energy Consumption per Successful Packet (SF9).]{
		\includegraphics[width=0.45\linewidth]{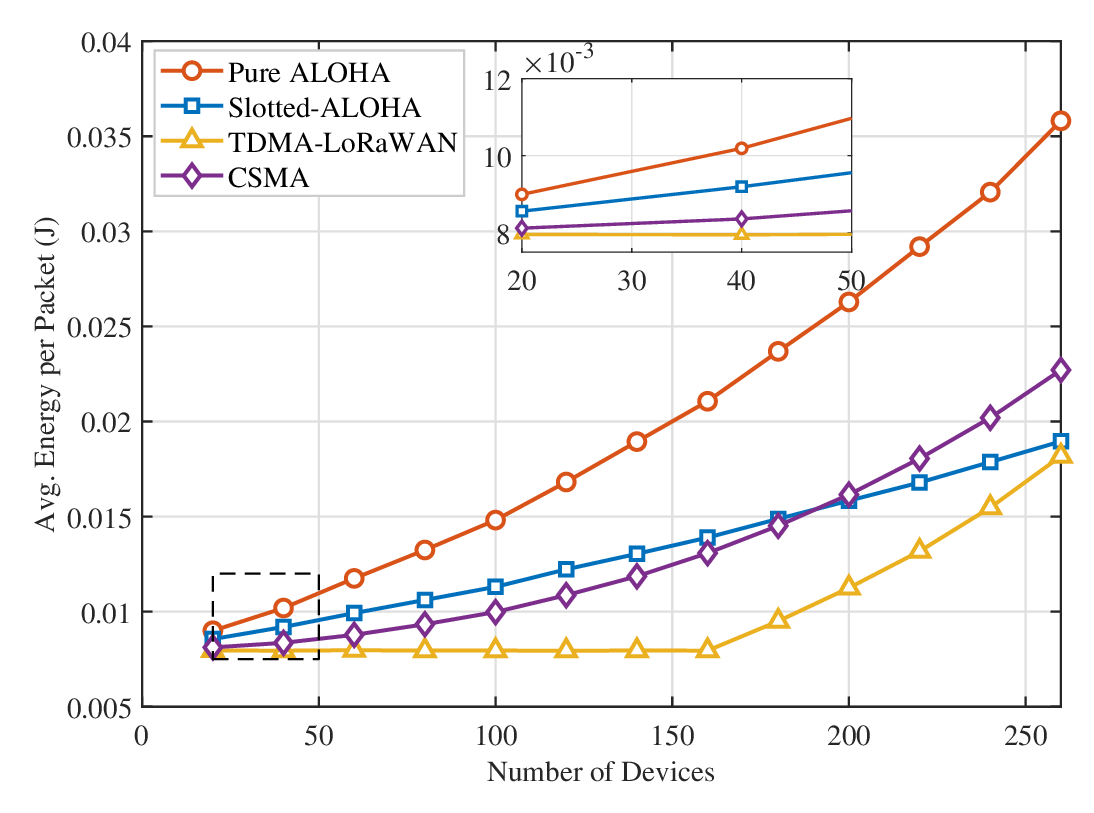}
		\label{fig:sf9_consumption}
	}
	\caption{Comparison of Average Energy Consumption per Successful Packet.}
	\label{Average_Energy_Consumption}
\end{figure*}

	\subsubsection{Success Ratio and Throughput Analysis}

	We evaluated the PDR and system throughput as the number of devices increased. Figs.~\ref{success_comparison} and~\ref{throughput_comparison} illustrate the comparative PDR and throughput results under both SF7 and SF9 settings. Note that the x-axis scales differ between scenarios due to the varying channel capacities inherent to different SFs.
	
	Under the SF7 scenario (Fig.~\ref{success_comparison}(a)), where the airtime is short, the collision probability is relatively low for small networks. Consequently, S-ALOHA and CSMA protocols perform moderately well. However, as the network density increases, the standard LoRaWAN performance degrades rapidly. While the alternative optimization schemes demonstrate a noticeable PDR increment over the baseline, TDMA-LoRaWAN significantly outperforms them. In contrast, the proposed TDMA-LoRaWAN maintains a high success rate by suppressing intra-network collisions through strict time-slot scheduling.
		
	The superiority of the proposed method is more pronounced in the SF9 scenario (Fig.~\ref{success_comparison}(b)). Due to the extended ToA, the channel saturates quickly with fewer devices. While the performance of ALOHA, S-ALOHA, and CSMA deteriorates rapidly as the number of devices increases, TDMA-LoRaWAN exhibits stronger robustness. It sustains a stable success rate and maintains high channel utilization even under high-load conditions. However, when the number of devices exceeds the total limit of assignable time slots in the TDMA system, the performance degrades rapidly. Nevertheless, it remains significantly superior to the other access protocols.
	
	The throughput analysis presented in Fig.~\ref{throughput_comparison} further shows that TDMA-LoRaWAN achieves higher throughput than the contention-based protocols across the evaluated scenarios, indicating a clear throughput advantage of the proposed scheduling approach.
	
	\subsubsection{Energy Efficiency Evaluation}

	Energy consumption is a key metric for battery-constrained IoT devices. In this study, we evaluate the average energy per successful packet to quantify the energy efficiency of different protocols under varying network densities. To ensure a rigorous assessment, the energy model for TDMA-LoRaWAN and Slotted ALOHA explicitly incorporates the overhead of periodic OOB synchronization.
		
	The total energy consumption $E_{\text{total}}$ is modeled as the sum of energy consumed in four primary components:
	\begin{equation}
		E_{\text{total}} = E_{\text{tx}} + E_{\text{rx}} + E_{\text{sync}} + E_{\text{sleep}},
		\label{eq:total_energy}
	\end{equation}
	where $E_{\text{tx}}$ and $E_{\text{rx}}$ denote the aggregate energy consumed by all data transmission and reception events (including retransmissions and receive windows), respectively. $E_{\text{sync}}$ represents the total energy overhead for synchronization maintenance, and $E_{\text{sleep}}$ is the baseline energy consumption in the sleep state.
	
	We model synchronization maintenance in the steady state. The synchronization interval is configured as an integer multiple of the beacon broadcast period, ensuring temporal alignment between device wake-up and beacon transmission. To compensate for accumulated clock drift, we employ a fixed listening window of $T_{\text{listen}} = 200$~ms centered around the expected arrival time. This conservative window provides a sufficient safety margin to reliably capture the beacon without requiring continuous channel monitoring. Since the energy consumed during frequency switching is negligible, the total synchronization energy is calculated as:
	\begin{equation}
		E_{\text{sync}} = N_{\text{sync}} \cdot P_{\text{rx}} \cdot T_{\text{listen}},
		\label{eq:sync_energy}
	\end{equation}
	where $N_{\text{sync}}$ is the number of synchronization beacons processed and $P_{\text{rx}}$ is the receiver power consumption.
		
	As illustrated in Fig.~\ref{Average_Energy_Consumption}, although OOB synchronization introduces a periodic baseline energy cost, TDMA-LoRaWAN still achieves the lowest average energy consumption per successful packet in dense scenarios. This is primarily because deterministic scheduling substantially reduces collision-induced energy waste and unnecessary retransmissions, which are the dominant sources of energy wastage in pure ALOHA and CSMA-based networks as device density increases.

	\subsubsection{Scalability Analysis}		
\begin{figure}[!t]
		\centering
		\includegraphics[width=2.95in]{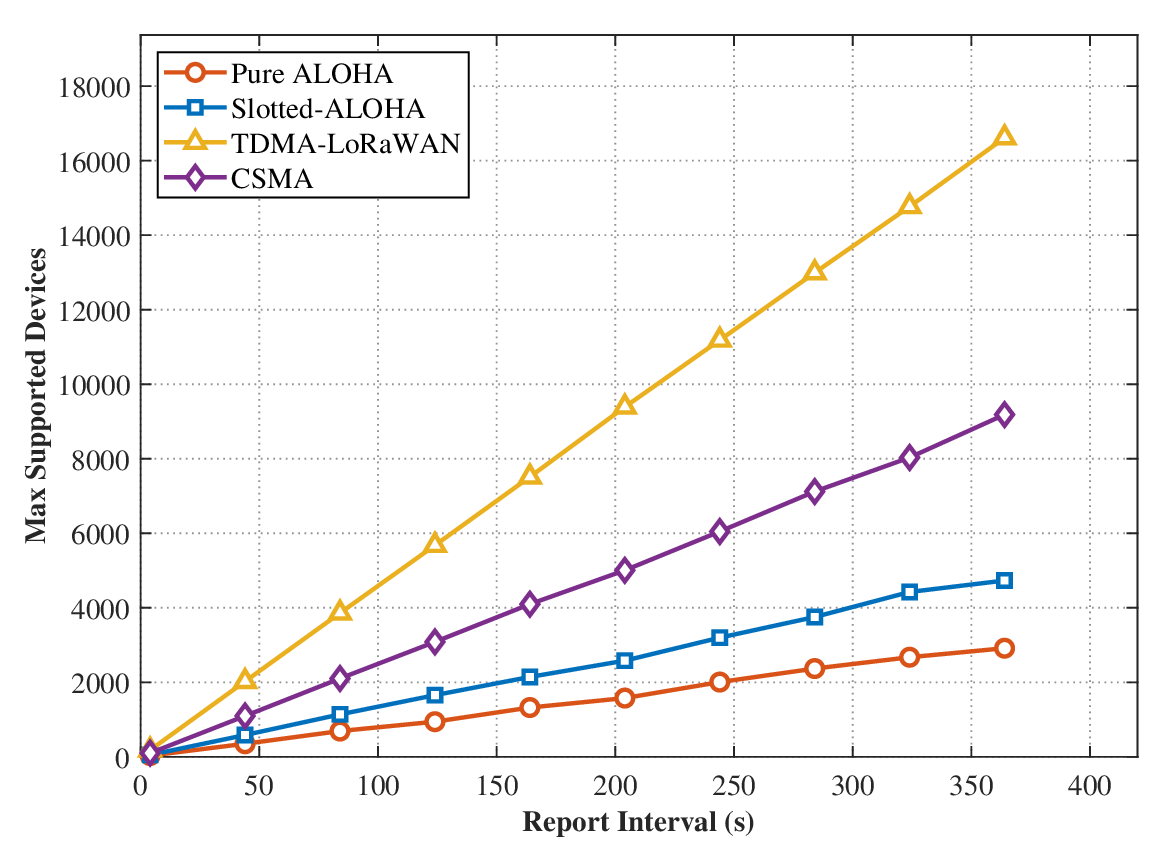}
		\caption{Maximum number of devices supported by the system.}
		\label{capacity}
	\end{figure}

	To quantify network scalability, we evaluated the maximum number of end devices supported by each protocol across different uplink cycle settings. With the SF fixed at SF9, Fig.~\ref{capacity} delineates the capacity boundaries as the uplink cycle extends from 4~s to 400~s, subject to a PDR threshold of 80\%. The results demonstrate that the TDMA-LoRaWAN protocol consistently outperforms its counterparts in terms of network capacity, primarily because scheduled resource allocation reduces random-access contention within the nominal operating capacity. While prolonged uplink cycles effectively alleviate channel congestion and augment system capacity, they inherently introduce higher data latency, necessitating a critical trade-off in system dimensioning.
	
	\section{Conclusion}
	This work presents a lightweight overlay TDMA scheme to alleviate the scalability and capacity limitations of LoRaWAN networks. By implementing an OOB synchronization mechanism and a flexible superframe structure, the proposed protocol supports heterogeneous device requirements while maintaining backward compatibility with standard Class~A end nodes. The design requires no hardware modifications and provides a practical low-cost enhancement path for IoT applications with dense uplink reporting demands.
	
	Experimental validation on a 20-node indoor positioning testbed demonstrated that the proposed solution significantly outperforms standard LoRaWAN under the evaluated reporting configuration. Specifically, the measured results showed a throughput increase of 30\% and a reduction in packet loss rate from 25.8\% to 5.02\%, confirming the benefit of scheduled channel access in the tested scenario. In addition, large-scale simulations corroborated these findings, supported the scalability analysis under larger network sizes, and indicated improved energy efficiency per successful packet in dense networks. Strict collision-free operation is guaranteed within the nominal resource capacity of the configured scheduler, whereas beyond this capacity limit the design preserves access for higher-priority traffic through controlled resource reuse.
	
	While the current static scheduling remains deterministic within this nominal capacity region, it may lack flexibility under highly bursty traffic conditions. Therefore, future work will focus on integrating adaptive time-slot allocation and more flexible resource reconfiguration mechanisms to better respond to dynamic network loads. In addition, we plan to extend the experimental validation to larger and more diverse indoor deployment scenarios.

	\FloatBarrier
	\footnotesize
	\bibliography{references}

\end{document}